\documentclass{article}
\usepackage[a4paper, margin=1in]{geometry}
\usepackage[T1]{fontenc}

\usepackage{color}
\usepackage{booktabs}
\usepackage{comment}
\usepackage{amsmath}
\usepackage{tabularx}
\usepackage{hyperref}
\usepackage{authblk}
\usepackage{subcaption}
\usepackage{subcaption}
\usepackage{graphicx}
\usepackage[table,xcdraw]{xcolor}
\usepackage{tcolorbox}
\usepackage[skip=0.7em]{parskip}
\usepackage{float}

\usepackage{algorithmic}
\usepackage[ruled,vlined]{algorithm2e}
\hypersetup{
	bookmarks=true,         
	unicode=false,          
	pdftoolbar=true,        
	pdfmenubar=true,        
	pdffitwindow=false,     
	pdfstartview={FitH},    
	pdftitle={Do LLMs Free-Ride? behavioural Intention Recognition in Multi-Agent Public Goods Games},    
	pdfauthor={Author},     
	pdfsubject={Subject},   
	pdfcreator={Creator},   
	pdfproducer={Producer}, 
	pdfkeywords={keywords}, 
	pdfnewwindow=true,      
	colorlinks=true,        
	linkcolor=blue,         
	citecolor=blue,         
	filecolor=magenta,      
	urlcolor=cyan           
}

\title{
 \textbf{Understanding LLM Agent Behaviours via Game Theory: Strategy Recognition, Biases and Multi-Agent Dynamics}
}

\author[2,3,$\dagger$]{Trung-Kiet Huynh}
\author[2,3,$\dagger$]{Duy-Minh Dao-Sy}

\author[1,3]{Thanh-Bang Cao}
\author[1,3]{Phong-Hao Le}
\author[1,3]{Hong-Dan Nguyen}
\author[2,3]{Phu-Quy Nguyen-Lam}
\author[1,3]{Minh-Luan Nguyen-Vo}
\author[1,3]{Hong-Phat Pham}
\author[2,3]{Phu-Hoa Pham}
\author[1,3]{Thien-Kim Than}
\author[2,3]{Chi-Nguyen Tran}
\author[1,3]{Huy Tran}
\author[1,3]{Gia-Thoai Tran-Le}

\author[4]{Alessio Buscemi}
\author[1,3,$\star$]{Le Hong Trang}
\author[5,$\star$]{The Anh Han }

\affil[1]{Faculty of Computer Science and Engineering, Ho Chi Minh City University of
Technology (HCMUT), Vietnam}
\affil[2]{Faculty of Information and Technology, Ho Chi Minh City University of Science (HCMUS), Vietnam}
\affil[3]{Vietnam National University - Ho Chi Minh City (VNU-HCM), Vietnam}
\affil[4]{Luxembourg Institute of Science and Technology, Esch-sur-Alzette, Luxembourg}
\affil[5]{School of Computing, Engineering and Digital Technologies, Teesside University, United Kingdom}
\affil[$\dagger$]{Equal Contribution}
\affil[$\star$]{Corresponding authors: The Anh Han (Email: t.han@tees.ac.uk), Le Hong Trang (Email: lhtrang@hcmut.edu.vn)}

\date{ }

\begin{document}

\maketitle

\begin{abstract}
As Large Language Models (LLMs) increasingly operate as autonomous decision-makers in interactive and multi-agent systems and human societies, understanding their strategic behaviour has profound implications for safety, coordination, and the design of AI-driven social and economic infrastructures. Assessing such behaviour requires methods that capture not only what LLMs output, but the underlying intentions that guide their decisions. In this work, we extend the FAIRGAME framework to systematically evaluate LLM behaviour in repeated social dilemmas through two complementary advances: a payoff-scaled Prisoner’s Dilemma isolating sensitivity to incentive magnitude, and an integrated multi-agent Public Goods Game with dynamic payoffs and multi-agent histories. These environments reveal consistent behavioural signatures across models and languages, including incentive-sensitive cooperation, cross-linguistic divergence and end-game alignment toward defection. To interpret these patterns, we train traditional supervised classification models on canonical repeated-game strategies and apply them to LLM decisions in FAIRGAME, showing that LLMs exhibit systematic, model- and language-dependent behavioural intentions, with linguistic framing at times exerting effects as strong as architectural differences. Together, these findings provide a unified methodological foundation for auditing LLMs as strategic agents and reveal systematic cooperation biases with direct implications for AI governance, collective decision-making, and the design of safe multi-agent systems.

\end{abstract}

\section{Introduction}
\label{sec:intro}

Large language models (LLMs) are increasingly deployed as \emph{agents} that interact with human users and with one another in recommendation systems, negotiation tools, and multi-agent assistants \cite{tessler2024ai, patel2020leveraging, hammond2025multi}. In these settings, LLMs are repeatedly exposed to \emph{cooperation dilemmas}, where they may produce behaviours that resemble contributing to a shared goal, free-riding on others, or enforcing social norms \cite{lu2024llms, tessler2024ai,balabanova2025media,commitment}. Such behaviour does not reflect genuine intentions or internal goals. It is the result of statistical patterns learned during training, combined with the incentives and context provided during interaction.
Evaluating these systems therefore requires more than verifying factual correctness or conversational quality. It requires analysing the \emph{emergent strategies} that LLM-based agents tend to exhibit over time, how these strategies are shaped by prompting, reward structures, and role assignment, and how they differ across languages and tasks \cite{buscemi2025llms, correia2025alife, akata2025, jia2025large, sun2025game, mao2025alympics, proverbio2025can, buscemi2025strategic}.

We adopt the notion of ``behavioural intention'' from the literature on repeated games, where an intention is operationalised as a decision rule that maps past interaction histories to current actions. Classical and evolutionary  game theory has long studied how such strategies emerge, stabilise, and can be inferred in repeated social dilemmas \cite{diStefanoIntention2023,Han2012ALIFEjournal}. Recent work builds on this by showing that canonical strategies---such as Always Cooperate (ALLC), Always Defect (ALLD), Tit-for-Tat (TFT), and Win-Stay-Lose-Shift (WSLS) \cite{Axelrod1980Effective,sigmund:2010bo}---can be recognised in the Iterated Prisoner’s Dilemma (IPD) by training classifiers on noisy trajectories of play between artificial agents \cite{diStefanoIntention2023}. This line of research suggests that behavioural intention can be treated as a learnable object: given enough action trajectories, we may infer which strategy best explains an agent’s behaviour, even in the presence of execution noise.

In parallel, frameworks such as FAIRGAME (Framework for AI Agents Bias Recognition using Game Theory)~\cite{buscemi2025fairgame} have begun to systematically probe LLMs using repeated normal-form games. FAIRGAME provides a controlled environment in which LLM agents are prompted to play a range of matrix games under different payoff structures and experimental conditions, allowing researchers to measure fairness, cooperation, and other social biases across models, languages, and personalities. However, existing implementations mainly focus on symmetric two-player interactions with homogeneous players and relatively simple outcome measures (e.g., average cooperation rates). They leave open important questions about (i) how sensitive LLM agents are to changes in incentives even when the underlying game is fixed, and (ii) how they behave in richer group settings where multiple players interact over a shared public good and where coalitions or coordination patterns may emerge.

In this work, we bring these strands together and extend FAIRGAME along two complementary directions. From a game-theoretic perspective, we design \emph{two} families of repeated-game experiments for LLM agents. First, we introduce a payoff-scaling module for the Prisoner’s Dilemma that multiplies all entries of a fixed payoff matrix by a scalar factor, thereby manipulating the \emph{stakes} of cooperation while preserving the underlying strategic structure \cite{han_when_2021}. This allows us to ask whether LLM agents are sensitive to the absolute magnitude of payoffs, and whether such sensitivity depends on the model family or the language of interaction. Second, we develop a genuinely multi-agent extension based on a three-player Public Goods Game (PGG) \cite{sigmund:2010bo} with configurable multiplication factors, languages, and injected personalities. This multi-agent module enables us to study free-riding, coordination, and coalition-like behaviour in collective social dilemmas where payoffs depend on group-level contributions rather than pairwise interactions.

From a data-driven perspective, we then treat the \textit{trajectories} (i.e. sequences of actions by LLMs) generated by these experiments as input to a machine learning pipeline for behavioural intention recognition. Following the protocol of Di Stefano et al. \cite{diStefanoIntention2023}, we generate synthetic IPD trajectories for the four canonical strategies (ALLC, ALLD, TFT, WSLS) under execution noise and train a suite of classifiers, including Logistic Regression \cite{logit}, Random Forests, and Long Short-Term Memory (LSTM) networks \cite{lstm}, to identify which architectures are most robust to noisy play. The best-performing model is then used as an intent recogniser on FAIRGAME logs: we encode LLM game histories into state-action sequences and infer which canonical strategies-and which mixtures thereof-best explain the observed behaviour. This allows us to move beyond raw cooperation rates and examine how LLM agents’ latent intentions vary across models, languages, personalities, and roles (e.g., first-mover vs.\ second-mover).

Overall, this game-theoretic and data-driven approach allows us to address the following research questions:
\begin{itemize}
  \item \textbf{RQ1.} When the strategic structure of the Prisoner’s Dilemma is held fixed but all payoffs are uniformly scaled, do LLM agents systematically change their cooperative behaviour as the stakes increase or decrease, and how does this sensitivity vary across models and languages?
  \item \textbf{RQ2.} Can FAIRGAME-style evaluations be extended beyond symmetric two-player matrix games to more general multi-agent cooperation settings, such as Public Goods Games with heterogeneous incentives and personality prompts, and what patterns of cooperation, free-riding, and coordination do LLM agents exhibit in these environments?
  \item \textbf{RQ3.} To what extent can we predict and classify the behavioural intentions of LLM agents in repeated cooperation dilemmas using machine learning, and what systematic biases (across models, languages, personalities, and positional roles) emerge when we interpret their gameplay through canonical strategy classes?
\end{itemize}
The following sections develop these contributions in detail. Section~\ref{sec:methodology} introduces our extensions to the FAIRGAME framework, including the payoff-scaled Prisoner’s Dilemma, the multi-agent Public Goods Game, and the behavioural intent recognition pipeline. Section~\ref{sec:results} reports empirical results for all experimental conditions, and the final sections summarise our findings, discuss limitations, and outline directions for future work.

\section{Background}

\subsection{LLMs and Game Theory}

\subsection{FAIRGAME}
\label{Recall}

FAIRGAME (Framework for AI Agents Bias Recognition using Game Theory) \cite{buscemi2025fairgame} provides the foundational computational infrastructure upon which our work is built. The framework was originally introduced to support systematic, reproducible evaluations of LLMs through controlled multi-agent game-theoretic experiments. It offers a unified pipeline for defining games, orchestrating interactions among LLM agents, and analysing emergent behavioural patterns across languages, personalities, and strategic configurations.

At the core of FAIRGAME is a clear separation between \emph{declarative} game specification and \emph{procedural} execution. Experimental conditions are defined in a JSON configuration file that specifies, for each game, the payoff structure, available actions, horizon (number of rounds or stopping rule), set of LLM backends, languages, and agent-level options such as personality descriptions or information about the opponent. Prompt templates, written as natural-language skeletons, are provided separately for each language. At runtime, FAIRGAME combines the configuration and the appropriate template, injecting concrete details such as the current payoff matrix, round index, and history of past actions and payoffs. This separation allows the same game-theoretic design to be instantiated consistently across different models, languages, and framing variants.

The execution pipeline then turns these specifications into trajectories of play. Given a configuration, FAIRGAME enumerates the required game instances (e.g., all combinations of LLM backend, language, and personality condition) and simulates each one as a repeated normal-form game. In every round, the framework constructs a prompt for each agent that includes the game rules, any contextual information (such as personality hints), and the full public history of previous rounds. The LLM’s textual output is parsed into a discrete action, payoffs are computed according to the specified game, and the history is updated. The result is a structured log for each run, containing round-by-round actions and payoffs for all agents, which can be directly used for downstream quantitative analysis.

In its original formulation, FAIRGAME has been primarily applied to symmetric two-player matrix games, such as variants of the Prisoner’s Dilemma and coordination games, to compare the behaviour of different LLMs and languages under fixed payoff structures. In this work, we extend FAIRGAME along two complementary directions. First, we introduce a payoff-scaling module for the Prisoner’s Dilemma to investigate how sensitive LLM agents are to changes in the magnitude of incentives. Second, we develop a multi-agent extension that instantiates a multi-player Public Goods Game, generating richer trajectories of group interaction that we subsequently analyse using machine learning methods for strategy and intent recognition.

\section{Methodology}
\label{sec:methodology}

In this section we describe the methodology adopted to address the three research questions introduced in \ref{sec:intro}. 

\subsection{Game stakes: Prisoner's Dilemma payoff-scaling}
\label{GameImportance}

We first examine the sensitivity of LLM agents to the absolute magnitude of incentives in a dyadic setting. To this end, we use a repeated Prisoner's Dilemma in which only the numerical values of the payoffs are scaled, while the underlying strategic structure of the game is kept fixed.  In this way, the ``stakes'' of the interaction are varied without changing best responses or the ranking of outcomes. It has been shown that, more cooperative behaviours are strongly influenced by this factor in the context of repeated  games \cite{han_when_2021}.

The row player's baseline payoff matrix for the Prisoner's Dilemma is
\[
\begin{array}{c|cc}
 & \text{Option A} & \text{Option B} \\
\hline
\text{Option A} & (6, 6) & (0, 10) \\
\text{Option B} & (10, 0) & (2, 2)
\end{array}
\]
where the first and second entries in each cell denote the payoffs to the row and column player, respectively. This matrix satisfies the standard Prisoner's Dilemma ordering $T(10) > R(6) > P(2) > S(0)$. To manipulate the stakes of the game without altering its strategic structure, we introduce a scalar parameter $\lambda > 0$ and multiply all payoffs by $\lambda$. In our experiments we consider three values
\[
\lambda \in \{0.1, 1.0, 10.0\},
\]
corresponding to attenuated, baseline, and amplified payoff magnitudes. For example, when $\lambda = 0.1$ the row player's payoff matrix becomes
\[
\begin{array}{c|cc}
 & \text{Option A} & \text{Option B} \\
\hline
\text{Option A} & (0.6, 0.6) & (1.0, 0) \\
\text{Option B} & (0, 1.0) &  (0.2, 0.2)
\end{array}
\]
and when $\lambda = 10.0$ it becomes
\[
\begin{array}{c|cc}
 & \text{Option A} & \text{Option B} \\
\hline
\text{Option A} & (60, 60) & (0, 100) \\
\text{Option B} & (100, 0) & (20, 20)
\end{array}
\]
with the ordering $T > R > P > S$ preserved in all cases. This construction isolates the effect of payoff magnitude while keeping the underlying game-theoretic incentives unchanged.

Two-player games between LLM agents are run using FAIRGAME as the simulation engine. Each game is played for a fixed, finite horizon of $T = 10$ rounds, and in every round both agents observe the full public history of past actions and payoffs before choosing their next move. Agents do not communicate outside of their action choices. For each parameter configuration, we simulate multiple independent runs to account for the stochasticity of LLM outputs. We evaluate three LLM backends: GPT-4o \cite{gpt4o}, Claude~3.5~Haiku \cite{claude3}, and Mistral~Large \cite{mistral}. To probe potential cross-lingual effects, the same game is instantiated in English and Vietnamese. In all conditions, neutral framing is employed: ``Option~A'' corresponds to defection and ``Option~B'' to cooperation, and the prompt does not contain any explicit moral or normative language. We test and evaluate the results of 40 games with 400 decisions per setting.

\subsubsection{Public Goods Game}
\label{Multigame}

The multi-agent setting is modelled as a repeated Public Goods Game (PGG) with group size $N = 3$, the smallest group in which non-dyadic effects such as coalition-like behaviour and conditional cooperation can arise. In each round $t$, agent $i \in \{1,2,3\}$ chooses whether to contribute a fixed amount $c = 10$ to a common pool or to keep this endowment. Let $s_{i,t} \in \{0,1\}$ denote whether agent $i$ contributes in round $t$, and let $\mathbf{s}_t = (s_{1,t}, s_{2,t}, s_{3,t})$ be the joint action profile. The total contribution is multiplied by a synergy factor $r$ and redistributed equally among all group members. The per-round payoff of agent $i$ is
\begin{equation}
    \pi_{i,t} = \frac{r \times \sum_{j=1}^{N} (s_{j,t} \cdot c)}{N} - (s_{i,t} \cdot c),
    \label{eq:payoff}
\end{equation}
which induces the standard public goods social dilemma: collective welfare is maximised when all agents contribute, but unilateral defection strictly increases individual payoff and enables free-riding.

To support this group interaction within FAIRGAME, the original two-player matrix game implementation is extended along three axes. First, the static $2 \times 2$ payoff matrix is replaced by a dynamic public goods payoff module, which computes the vector of payoffs $\boldsymbol{\pi}_t$ from the joint action profile $\mathbf{s}_t$ according to Eq.~\eqref{eq:payoff}. Second, the game history is generalised from bilateral outcomes to vector-valued records: for each round, FAIRGAME now stores the full tuple of actions and payoffs for all $N$ agents, enabling strategy analysis at the group level. Third, the prompt-generation mechanism is adapted so that LLM agents reason about multi-agent histories and group-level incentives rather than dyadic exchanges. public goods-specific templates present the rules, worked payoff examples derived from $(c,r,N)$, the current round index, and the full multi-agent history, and are instantiated separately for each agent.

The resulting execution loop preserves FAIRGAME’s overall control flow but operates over joint strategy profiles and dynamic payoffs. Algorithm~\ref{alg:pgg_main} summarises this extension of FAIRGAME. 

\begin{algorithm}[H]
\caption{Multi-agent PGG execution (extension of FAIRGAME Alg.~2--3)}
\label{alg:pgg_main}
\KwIn{Set of instantiated games $\mathcal{G}$}
\KwOut{Set of game histories $\mathcal{O}$}

$\mathcal{O} \gets \emptyset$\;

\ForEach{game $G \in \mathcal{G}$}{
    $t \gets 1$\;
    initialise empty history $\mathcal{H}$\tcp*[r]{as in FAIRGAME}

    \While{$t \leq G.\text{n\_rounds}$ and not\_met($G.\text{stop\_cond}$)}{
        $\mathbf{s}_t \gets \textbf{QueryAllAgents}(G, t)$\tcp*[r]{joint strategy profile (multi-agent)}
        $\boldsymbol{\pi}_t \gets \textbf{PGGPayoff}(\mathbf{s}_t, G.\text{params})$\tcp*[r]{dynamic public goods payoff}
        update agent scores with $\boldsymbol{\pi}_t$\;
        append $(\mathbf{s}_t, \boldsymbol{\pi}_t)$ to $\mathcal{H}$\tcp*[r]{vector-valued history}
        $t \gets t + 1$\;
    }

    $\mathcal{O} \gets \mathcal{O} \cup \{\mathcal{H}\}$\;
}
\Return{$\mathcal{O}$}\;
\end{algorithm}

Here \textbf{QueryAllAgents} generalises FAIRGAME's per-round interaction to the multi-agent PGG setting: for each agent in $G$, a PGG-specific prompt is constructed using the current round index, game parameters $(c,r,N)$, and the full multi-agent history; the underlying LLM is queried; and the textual response is parsed into a discrete action in $\{Contribute, Keep\}$. The procedure \textbf{PGGPayoff} applies Eq.~\eqref{eq:payoff} to the resulting joint strategy profile $\mathbf{s}_t$ to produce the payoff vector $\boldsymbol{\pi}_t$. Full pseudocode for game instantiation, the execution loop, and the per-round decision routine is provided in Appendix~\ref{app:pgg_impl}.

The experimental configuration systematically varies the incentive structure, language, and model family. Games are played for a fixed horizon of $T = 10$ rounds, and agents are informed of this horizon in advance. For each condition, three LLM-based agents (one per player) are instantiated, the public goods payoff module defined by Eq.~\eqref{eq:payoff} is attached, and prompts are generated via the PGG template described above. In every round, all agents are queried with their respective prompts, their textual responses are parsed into actions in $\{\text{Contribute}, \text{Keep}\}$, and the resulting joint action profile and payoff vector are appended to the game history.

Table~\ref{tab:experimental_config} summarises the parameters used in all PGG experiments. Each unique combination of language (English or Vietnamese), LLM backend, and multiplication factor ($r \in \{1.1, 2.0, 2.9\}$) is repeated ten times to account for stochasticity in model outputs and to obtain stable estimates of cooperation rates.

\begin{table}[h]
\centering
\caption{Configuration of the Public Goods Game experiments.}
\label{tab:experimental_config}
\begin{tabularx}{\linewidth}{@{}l l X@{}}
\toprule
\textbf{Parameter} & \textbf{Value} & \textbf{Description} \\ \midrule
Group size ($N$) & $3$ & Minimal group size to observe non-dyadic interactions and emergent coalition dynamics. \\ \addlinespace
Contribution cost ($c$) & $10$ & Fixed amount contributed to the common pool if the agent chooses to cooperate. \\ \addlinespace
Rounds ($T$) & $10$ & Fixed and known game length; used to probe potential end-game effects. \\ \addlinespace
Multiplication factor ($r$) & $1.1, 2.0, 2.9$ & Three incentive regimes, from weak to relatively strong gains from cooperation. \\ \addlinespace
Languages & EN, VN & Games instantiated in English and Vietnamese to probe cross-lingual differences. \\ \addlinespace
Runs per configuration & $10$ & Independent repetitions for each (LLM, $M$, language) condition. \\ \bottomrule
\end{tabularx}
\end{table}

\subsection{Machine Learning approaches for understanding LLM behaviour }

While the FAIRGAME framework provides complete gameplay trajectories and descriptive metrics such as cooperation rates and payoff sensitivities, these primarily capture what agents do rather than why they behave that way. Our goal is to uncover the latent behavioural intentions embedded within these decision sequences to better interpret the motivations behind agents’ actions and understand how LLM strategies differ from human strategies.

Prior to \cite{Han2012ALIFEjournal,diStefanoIntention2023}, the authors conducted experiments involving the generation and collection of large-scale gameplay data, enabling the inference and recognition of well-known strategies, while introducing varying levels of execution noise ($\epsilon$) to replicate the stochastic nature of LLM outputs. We aim to reproduce this approach to classify the underlying behavioural intentions exhibited during the LLMs’ gameplay turns. 

\begin{figure}[H]
    \centering
    \includegraphics[width=1\linewidth]{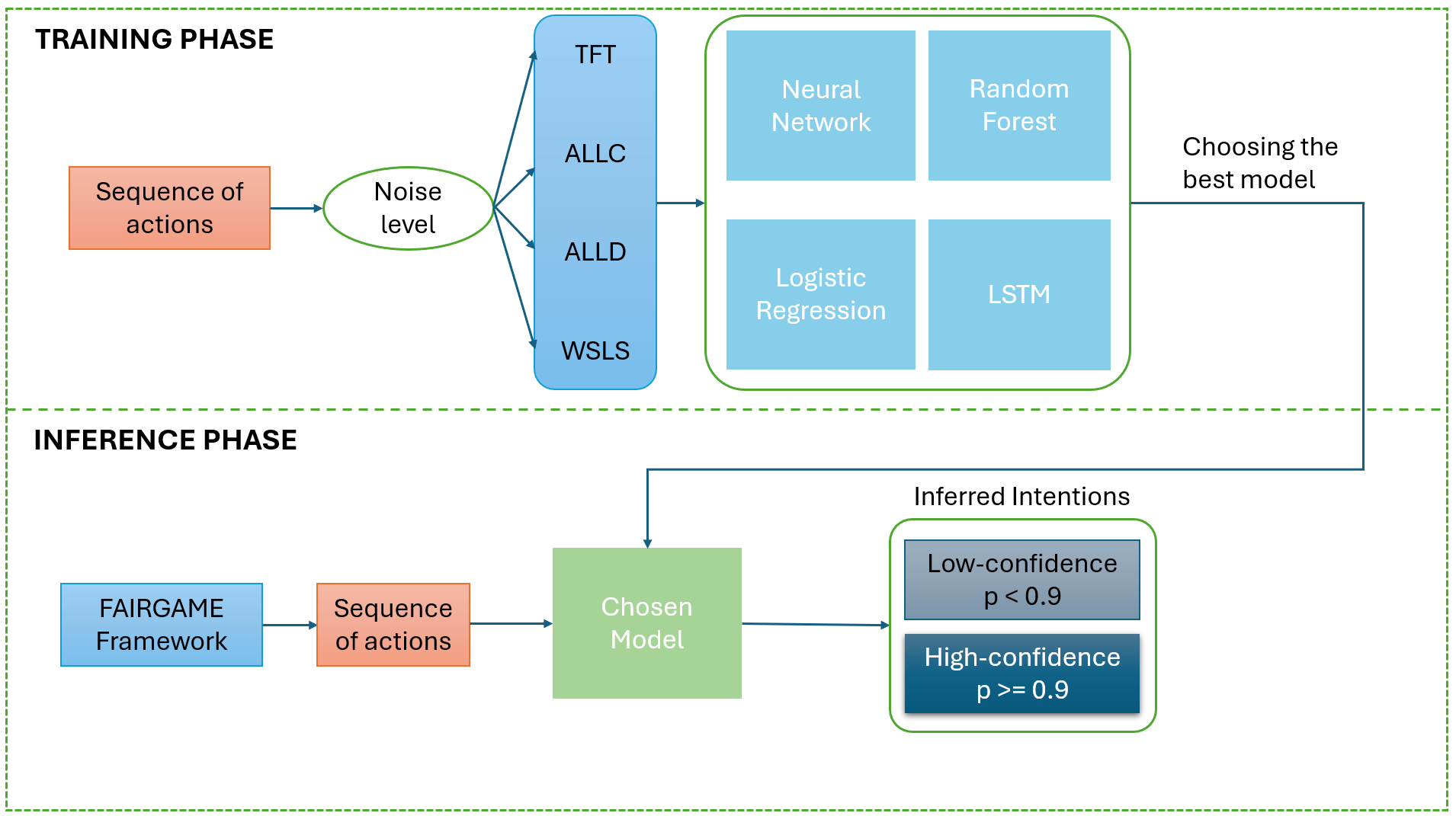}
    \caption{Machine Learning Pipeline for Understanding LLM Behaviour. Starting from action sequences associated with canonical strategies (ALLC, ALLD, TFT, GTFT, WSLS) under varying noise conditions, we train machine learning models to infer and classify underlying behavioural intentions. We then apply the best-performing model to the gameplay data generated by FAIRGAME. High-confidence predictions are used to identify which strategies the LLM adopts, whereas low-confidence cases are reserved for subsequent analysis to investigate the possibility of emerging behaviours by the LLM. }
    \label{fig:supervised_intention_model}
\end{figure}

Figure \ref{fig:supervised_intention_model} demonstrates the pipeline we employed to reproduce the behavioural intention prediction model and how we adapted this model to analyse the outputs of the FAIRGAME framework.

\label{ML}

\subsubsection{Training Phase}

Following the protocol and the synthetic interaction trajectories based on four canonical strategies: Tit-for-Tat (TFT), Always Cooperate (ALLC), Always Defect (ALLD), and Win-Stay-Lose-Shift (WSLS) defined in \cite{diStefanoIntention2023,sigmund:2010bo}. Crucially, to accommodate the stochastic imperfections inherent in generative AI, a noise parameter ($\epsilon \in \{0, 0.05\}$) is injected into these sequences. This robust synthetic dataset is then used to train and benchmark a suite of classifier architectures, including Logistic Regression \cite{cox1958regression}, Random Forests \cite{rf}, Neural Networks \cite{nn}, and Long Short-Term Memory (LSTM) \cite{lstm} networks. That allows us to identify the optimal model for capturing temporal strategic dependencies. The trained model is subsequently used to predict the strategies underlying the gameplay turns of LLM agents.

\subsubsection{Inference Phase}

The raw experimental data consists of simulation logs generated by the FAIRGAME framework. Each simulation run records the round-by-round interactions between a pair of agents, including their personalities, chosen actions, and resulting payoffs. To facilitate quantitative analysis, we preprocess these unstructured logs into a structured sequence format required for our intent recognition models. 

The FAIRGAME simulation logs record agent interactions using neutral labels: \texttt{OptionA} (representing Defection) and \texttt{OptionB} (representing Cooperation). Our encoding scheme transforms these raw sequences into a semantic "state-action" format that explicitly captures the conditional nature of game-theoretic strategies. We then determine the interaction outcome for each round $(t-1)$ based on the joint actions: \textit{Reward} ($R$, mutual cooperation), \textit{Punishment} ($P$, mutual defection), \textit{Temptation} ($T$, agent defects while opponent cooperates), and \textit{Sucker} ($S$, agent cooperates while opponent defects).

The chosen model is applied to the processed FAIRGAME trajectories. For each agent 
$i$ in a game session, the model outputs a probability distribution over the four strategy classes. To rigorously analyse the results and avoid over-interpreting ambiguous behaviours, we focus our downstream analysis on high-confidence predictions, defined as classifications where the model assigns a probability greater than 0.9 to a single strategy. Predictions with confidence below 0.9 require separate analysis to determine whether they reflect emerging LLM behaviours that differ from those of humans.

\section{Results}
\label{sec:results}

In this section, we present the results obtained using the methodology described in \ref{sec:methodology} to address the research questions targeted in this work.

\subsection{Payoff Magnitude Sensitivity in Prisoner's Dilemma}
\label{res:payoff_magnitude}
\begin{figure}[H]
    \centering
    \includegraphics[width=\linewidth]{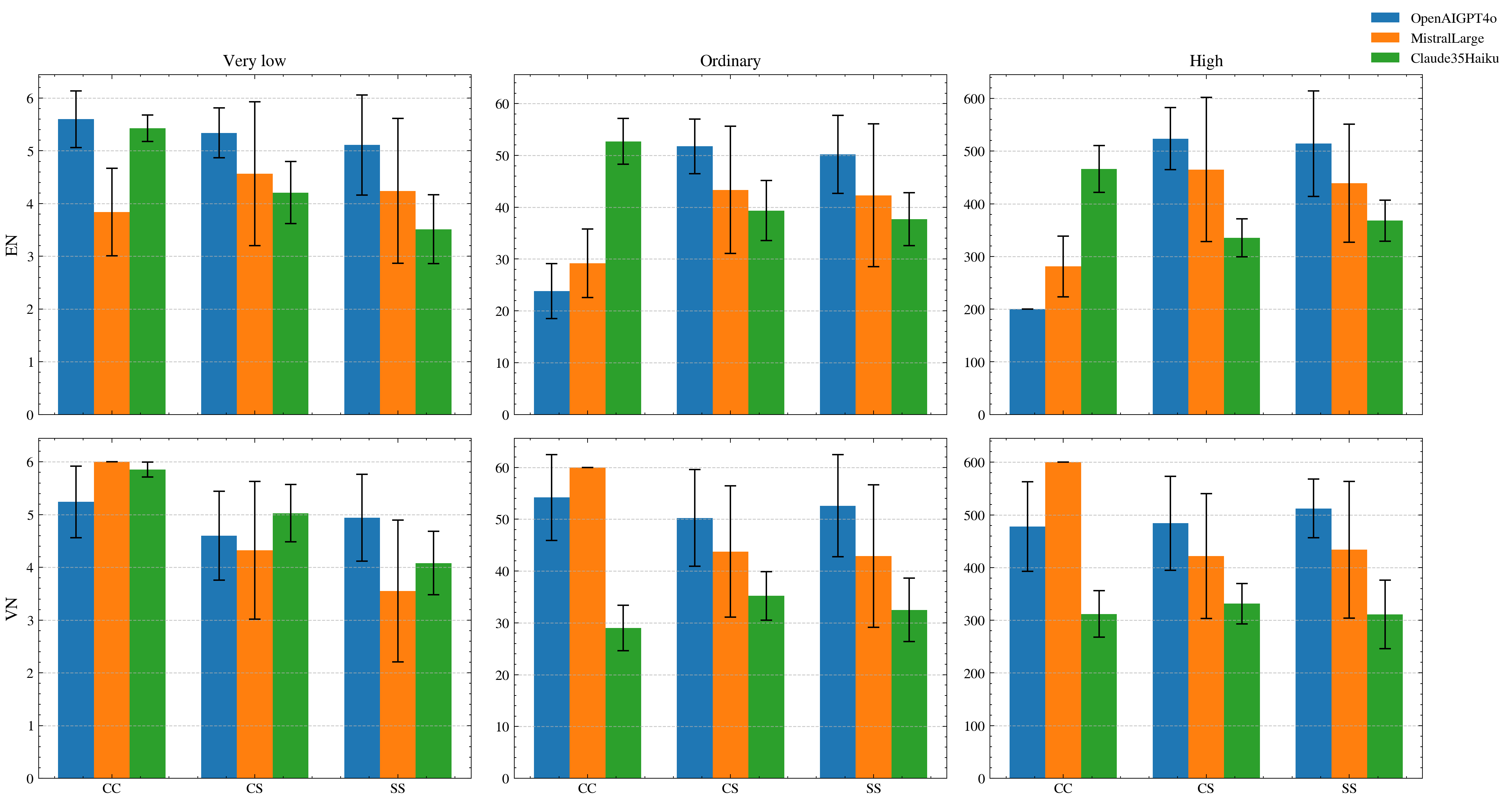}
    \caption{Aggregated final penalties across repeated Prisoner’s Dilemma games, presented for each LLM under different payoff magnitudes. Results are reported for both English (EN) and Vietnamese (VN), and evaluated across the scaling parameters $\lambda \in \{0.1, 1.0, 10.0\}$, which correspond to very low, ordinary, and high penalty scales, respectively.}
    \label{fig:3.1-total_penalties}
\end{figure}

Figure \ref{fig:3.1-total_penalties} displays bar plots summarizing the total penalties incurred by agents in the Prisoner’s Dilemma game, with 95\% Confidence Interval. These totals are computed from the test results and grouped by the scaling parameter $\lambda$, language, and personality pairings. In the default FAIRGAME configuration, each agent is assigned one of two personalities: Cooperative (C) or Selfish (S), resulting in three possible pairings: mixed (CS), both selfish (SS), and both cooperative (CC).
Because the payoff matrix is scaled by $\lambda$, the range of total penalties scales accordingly. Figure \ref{fig:3.1-total_penalties} shows that the very low–magnitude payoff setting (i.e., $\lambda = 0.1$) consistently produces higher overall penalties across models, languages, and personality pairings. This indicates that when the stake of the game is very low,  defection is highly frequent. This is in line with the game theory analysis in \cite{han_when_2021} (see Figure 4 therein). Within each language, the ordinary and high payoff settings yield broadly similar patterns, with only minor variations at specific points.

When comparing languages, the results indicate that several LLMs are sensitive to the linguistic context. Cooperative pairings in the Vietnamese context strongly favour defection, considering the two agents GPT-4o and Mistral. Notably, when shifting from English to Vietnamese, LLM agents often reverse their behaviour: models that yield lower total penalties in English tend to produce higher penalties in Vietnamese, and vice versa.

\begin{figure}[H]
    \centering
    \includegraphics[width=\linewidth]{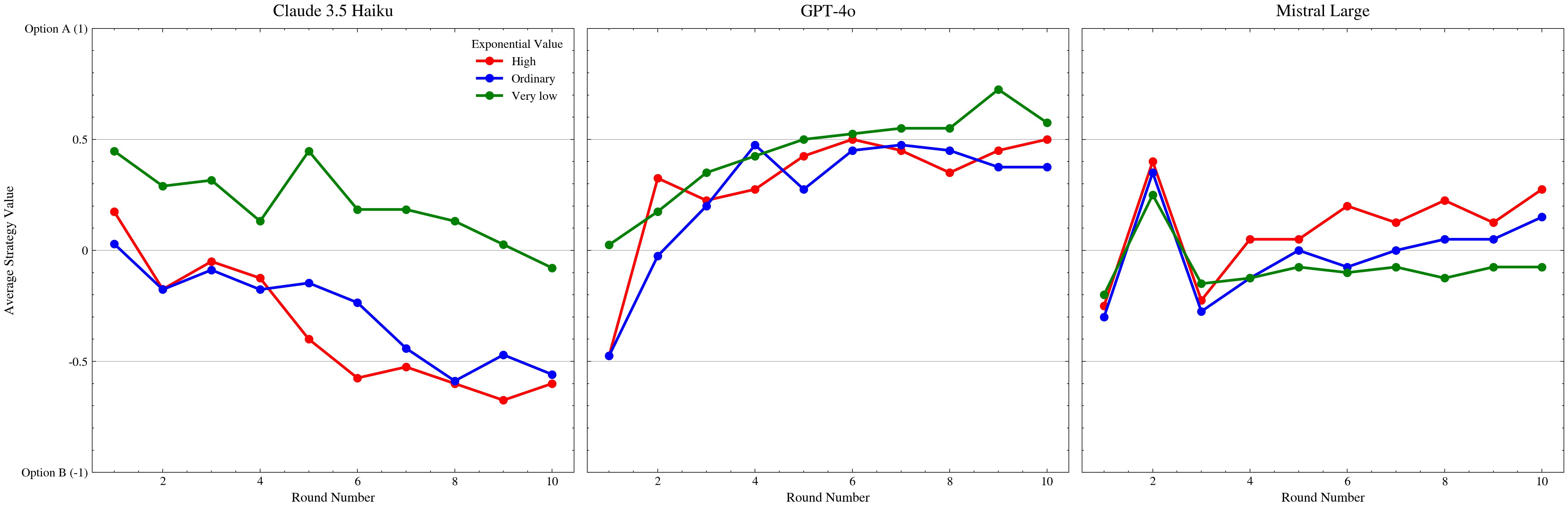}
    \caption{Average trajectory of strategy choices across repeated rounds in all Prisoner’s Dilemma experiments, shown for each LLM under different payoff magnitudes. A value of $1$ indicates selection of Option A (defection), while $-1$ corresponds to Option B (cooperation). The experiments consider $\lambda \in \{0.1, 1.0, 10.0\}$, representing high, ordinary, and very low penalty scales, respectively. The \textcolor{blue}{blue} line denotes the standard payoff matrix ($\lambda = 1.0$), the \textcolor{red}{red} line reflects the payoff matrix scaled by $10$ ($\lambda = 10.0$), and the \textcolor{green}{green} line represents the payoff matrix scaled by $0.1$ ($\lambda = 0.1$).}
    \label{fig:3.1-po_sensitivity_results}
\end{figure} 
Figure \ref{fig:3.1-po_sensitivity_results} presents the sequences of choices  across rounds for each model under varying values of the scaling parameter $\lambda$. For Claude 3.5 Haiku, the relationship between payoff magnitude and strategic behaviour appears relatively weak and inconsistent \cite{razeghi2022impact}. GPT-4o, however, shows a clearer pattern, exhibiting increasingly selfish behaviour as the payoff matrix is scaled down. In contrast, Mistral Large demonstrates the opposite tendency, with its behaviour shifting counter to the trend observed in GPT-4o.

The performance trajectories of the Claude 3.5 Haiku model in Figure~\ref{fig:3.1-po_sensitivity_results} exhibit a general downward trend as the number of iterations increases. The very low trajectory, predominantly situated in the upper quadrant (favouring Option A), displays a gradual shift towards the lower quadrant (Option B), interrupted by a minor peak at Round 5 that suggests a notable degree of strategic volatility. Conversely, the high and ordinary trajectories are primarily concentrated in the lower domain, indicating a cooperative tendency designed to yield a higher average expected utility. The very low line demonstrates a significantly greater inclination towards the upper spectrum compared to the high line, a divergence that may be partly attributed to linguistic bias \cite{bang2023multitask}. Specifically, within the context of the Vietnamese language combined with a payoff scalar of $\lambda = 0.1$, Claude 3.5 Haiku appears to misinterpret the unit magnitude of the payoff matrix, resulting in the erroneous maximization of penalties rather than their intended minimization.

A similar pattern is observed in the GPT-4o model, where the very low trajectory exhibits a more pronounced bias towards the upper spectrum relative to the other two lines. 

Regarding the Mistral Large model, empirical observations indicate that higher payoff values correlate with an increased likelihood of defection, a behaviour consistent with the dominant strategy concept in game theory. Furthermore, a distinct visual correlation is evident, characterized by a substantial spike at Round 2. This anomaly suggests a retaliatory mechanism triggered by a defection in the initial round, or alternatively \cite{akata2025}, represents a probing tactic aimed at minimizing the maximum potential penalty.

\subsection{LLM behaviours in Public Goods Game}
\label{res:pgg}
We now turn to the multi-agent Public Goods Game to address RQ2: how do LLM agents behave in a  collective group social dilemma, and how does this behaviour depend on incentives, time, and framing (model, personality, language).

\subsubsection{Cooperation Rates Across Multiplication Factors}
\label{res:pgg_cooperation}

\begin{figure}[H]
    \centering
    \begin{subfigure}[b]{\textwidth}
        \centering
        \includegraphics[width=0.95\textwidth]{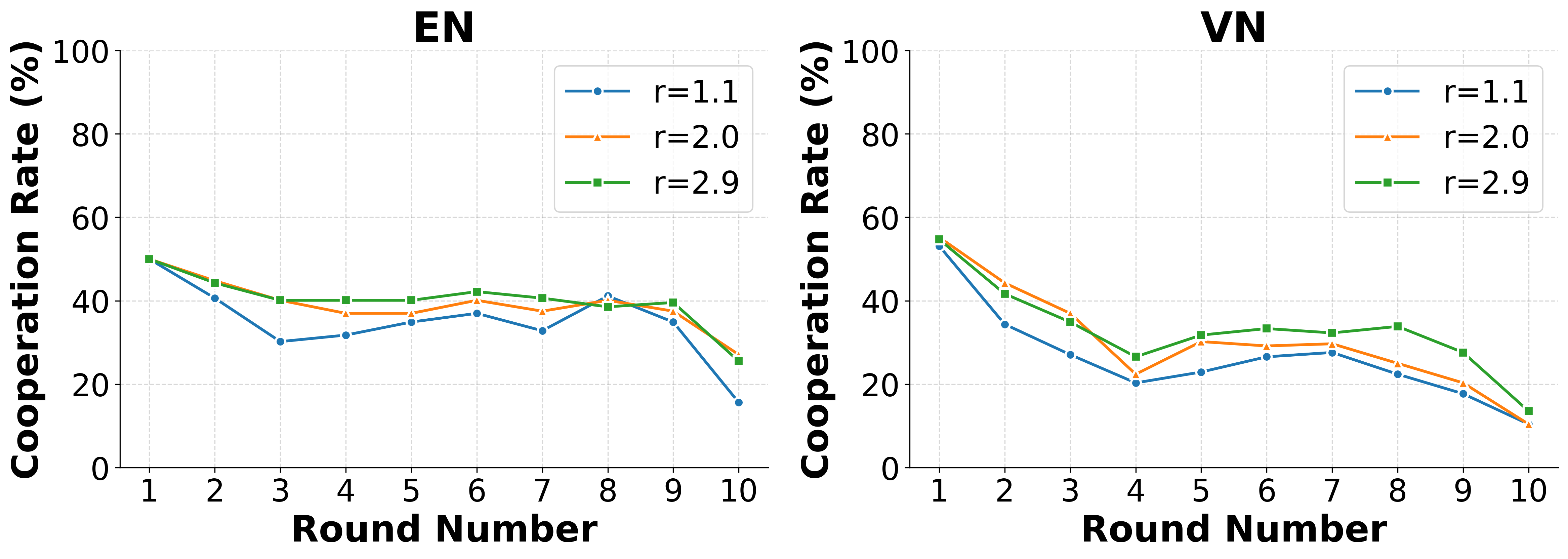}
        \caption{Claude 3.5 Haiku}
        \label{fig:claude_cooperation}
    \end{subfigure}
    
    \vspace{0.5cm}
    
    \begin{subfigure}[b]{\textwidth}
        \centering
        \includegraphics[width=0.95\textwidth]{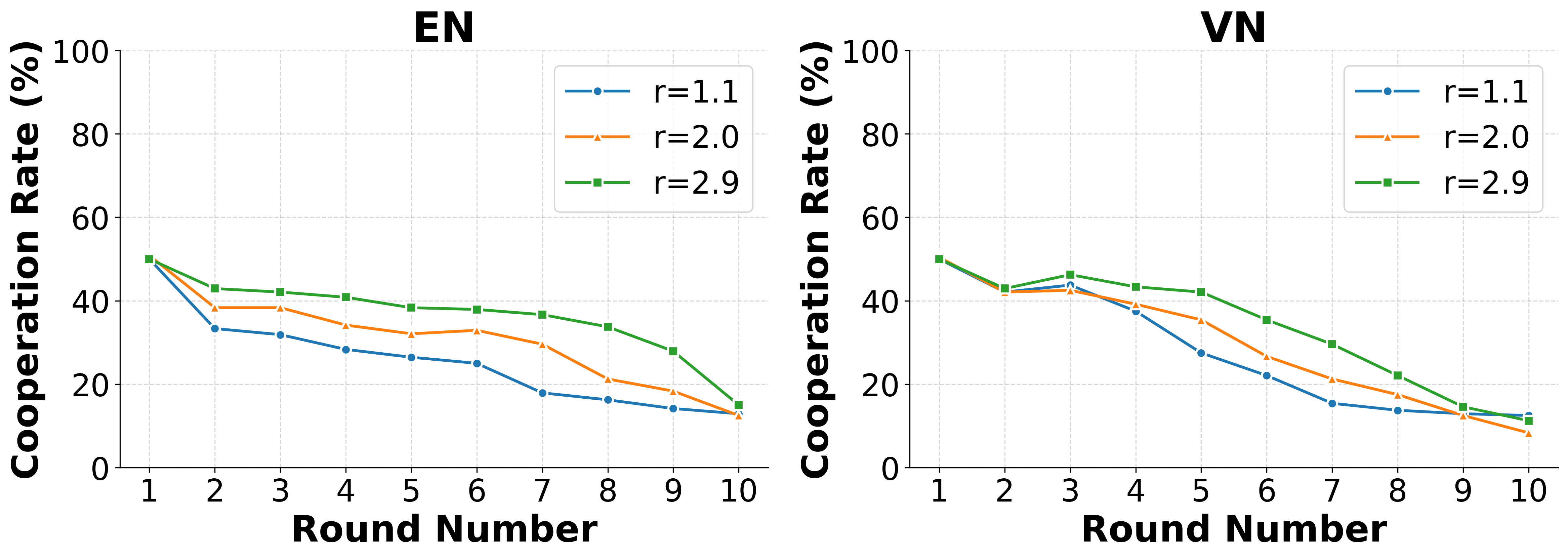}
        \caption{Mistral Large}
        \label{fig:mistral_cooperation}
    \end{subfigure}

    \vspace{0.5cm}

    \begin{subfigure}[b]{\textwidth}
        \centering
        \includegraphics[width=0.95\textwidth]{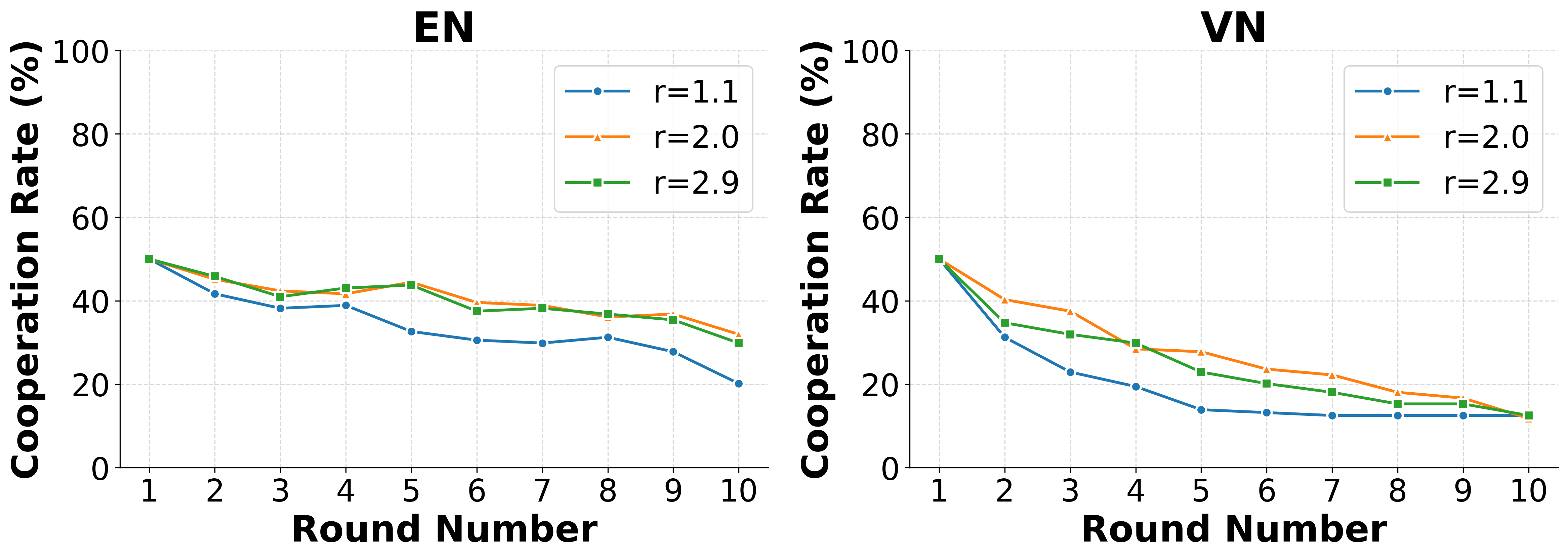}
        \caption{GPT-4o}
        \label{fig:mistral_cooperation}
    \end{subfigure}
    
    \caption{Comparison of cooperation rate evolution across 10 rounds for three LLM models under different multiplication factors ($r \in \{1.1, 2.0, 2.9\}$) and language conditions.}
    \label{fig:cooperation_comparison}
\end{figure}

Figure~\ref{fig:cooperation_comparison} presents the average round-by-round cooperation rates for Claude 3.5 Haiku (averaged over 8 independent runs), Mistral Large (10 independent runs) and ChatGPT 4.0 (6 different runs) under three multiplication factors ($r \in \{1.1, 2.0, 2.9\}$) and two languages (English on the left, Vietnamese on the right). Because each curve represents an average over repeated stochastic simulations, differences in smoothness, variability, and separation between curves reflect not only each model’s strategic tendencies but also its internal stability across repeated interactions.

Across all conditions, the overall level of cooperation increases  with the multiplication factor. This  is consistent  with game theoretical predictions in PGG \cite{sigmund:2010bo,van2012emergence}.    This observation  holds  for all LLM models (Claude vs. Mistral vs. ChatGPT) and languages (English vs. Vietnamese) for the majority of early, mid, and late rounds.

Despite this shared structure, the trajectories reveal systematic cross-linguistic differences. Cooperation in the English condition (left panels) tends to begin at higher values and decline more smoothly over time, whereas cooperation in the Vietnamese condition (right panels) often drops more steeply in the early rounds and falls to lower levels by the game’s conclusion. The fact that these differences persist even after averaging multiple independent runs suggests that they are not random fluctuations but reflect genuine differences in how the two linguistic framings shape the models’ interpretation of the task. English prompts may encode clearer normative cues or more strongly activate training-data priors related to fairness and reciprocity, whereas Vietnamese prompts may elicit more payoff-maximizing or conservative strategies. Such linguistic asymmetries are particularly visible in rounds 1-3: cooperation begins at roughly 40-60\% across conditions, but the Vietnamese trajectories collapse more rapidly, implying that early free-riding by one agent-inevitable in some fraction of stochastic runs-is interpreted more pessimistically by the models in this language.

\subsubsection{Cooperation Dynamics and End-Game Effects}
\label{res:pgg_temporal}

\begin{figure}[H]
    \centering
    \begin{subfigure}[b]{0.48\textwidth}
        \centering
        \includegraphics[width=\textwidth]{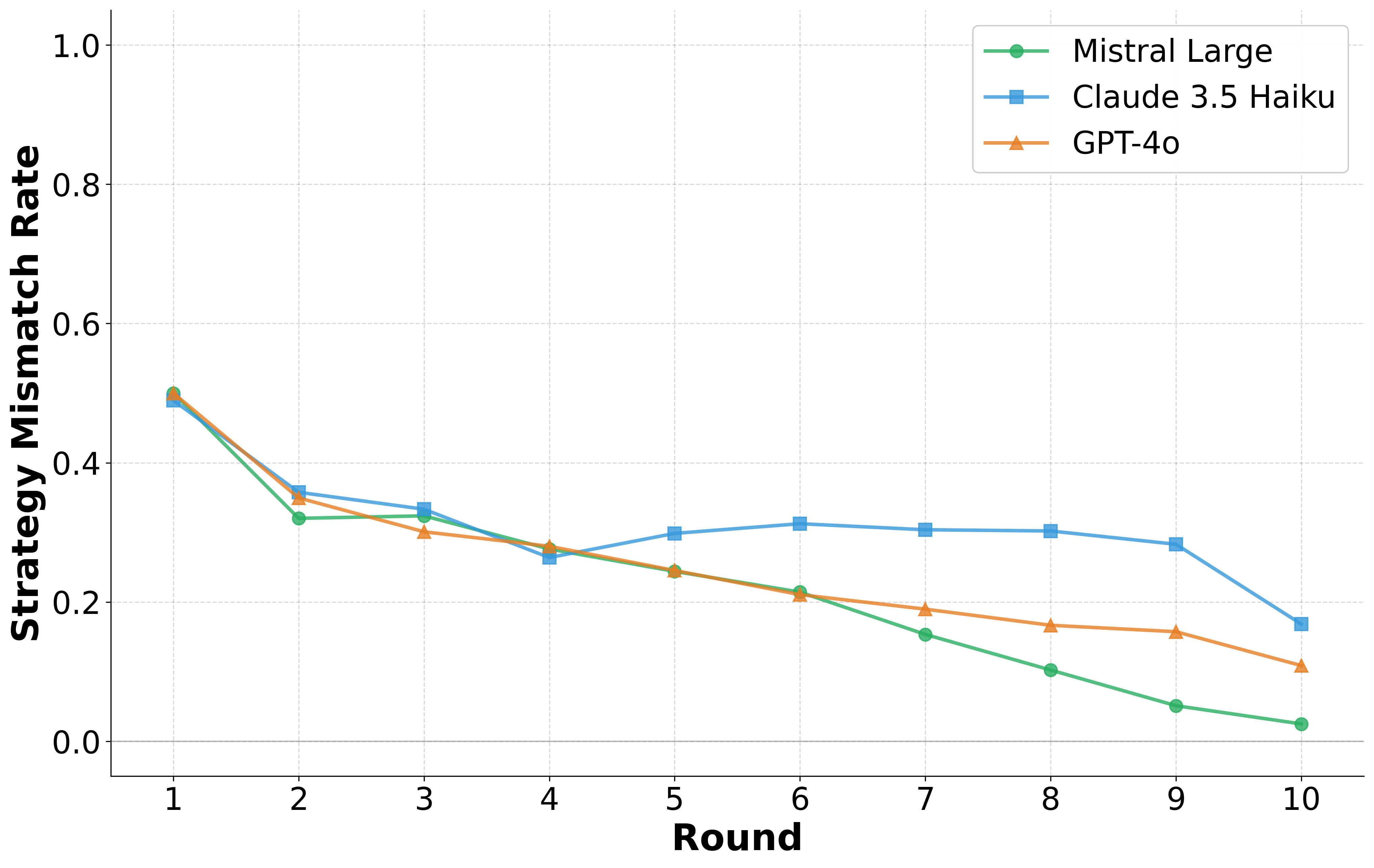}
        \caption{}
        \label{}
    \end{subfigure}
    \hfill
    \begin{subfigure}[b]{0.48\textwidth}
        \centering
        \includegraphics[width=\textwidth]{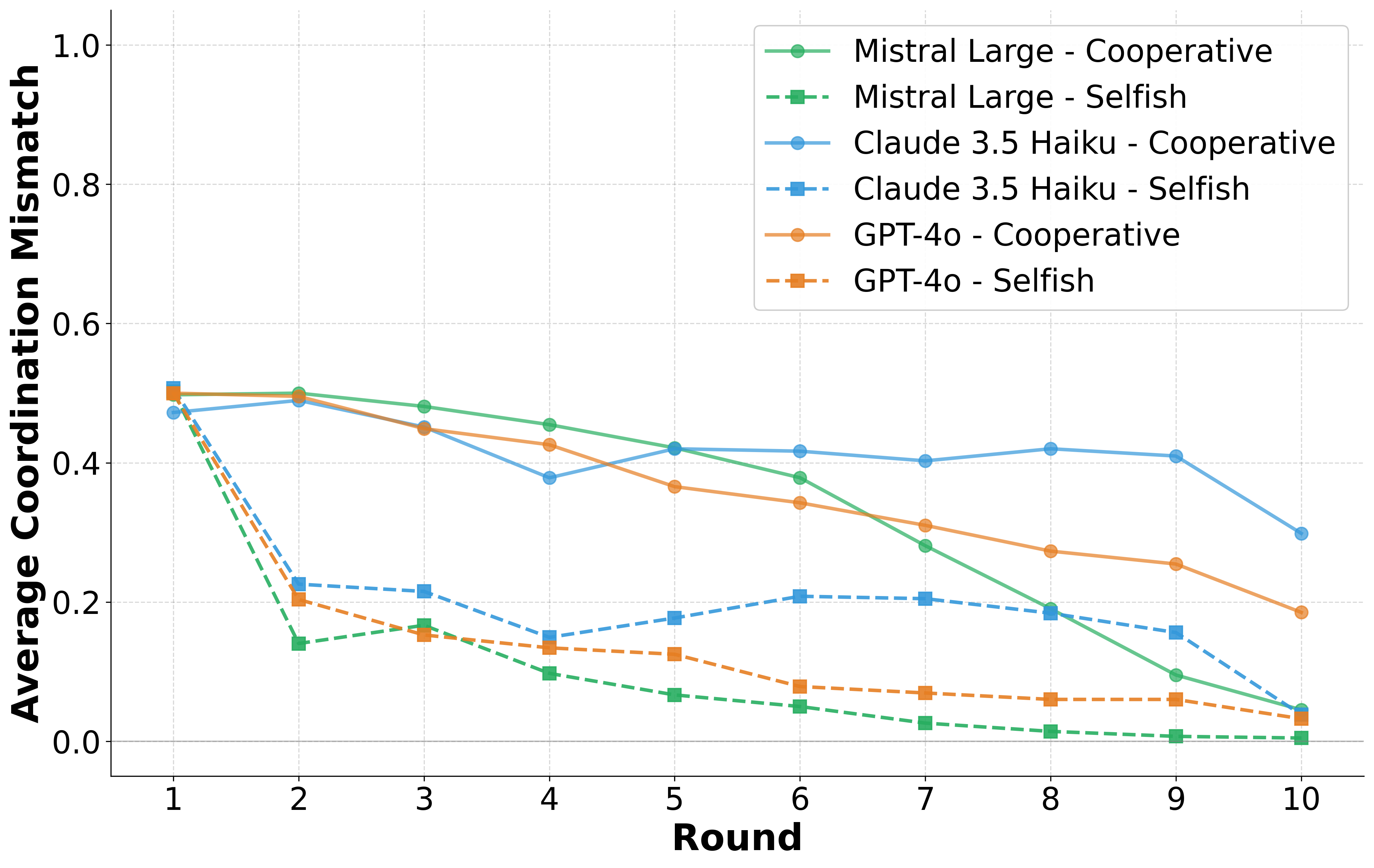}
        \caption{}
        \label{}
    \end{subfigure}
    
    \caption{Analysis of cooperation trends across the 10 rounds. Investigation of whether LLM agents exhibit end-game defection}
    \label{fig:pgg_temporal}
\end{figure}

Figure~\ref{fig:pgg_temporal} examines how the cooperation  evolves over the ten rounds and whether there is any special instability at the end of the game. Because mismatch is computed from whether all three agents select the same action (100\% match) or whether only two align ($\approx$ 66\% match), the curves quantify the degree to which agents act coherently versus independently. Both panels plot different notions of “mismatch”: panel (a) tracks how often agents’ strategies differ from one another (strategy mismatch rate), while panel (b) breaks this down further by personality condition (cooperative vs. selfish) and looks at average coordination mismatch.

The trajectories in panel (a) show a consistent pattern across all models. By the later rounds, they all demonstrate substantially lower mismatch, indicating that the agents converge toward a common behavioural mode. Mistral Large exhibits the strongest convergence, with mismatch approaching zero by round 10, whereas Claude 3.5 Haiku and GPT-4o maintain slightly higher but still reduced levels of disagreement. This means that although cooperation levels collapse in the final rounds (as shown earlier in Figure~\ref{fig:cooperation_comparison}), this collapse is not accompanied by chaotic or divergent play; instead, the agents converge together toward broadly similar non-cooperative choices. 

Panel (b) further decomposes these coordination dynamics by agent personality. Under selfish personality instructions, all three models rapidly settle into a mutually consistent pattern: mismatch drops sharply after the first round and approaches zero well at the end of the game, with very little variation across rounds. Under cooperative instructions, however, mismatch remains substantially higher and decreases only gradually. The cooperative condition thus sustains a wider diversity of behaviours: some agents continue contributing while others defect, even deep into the game. Among the three models, Claude 3.5 Haiku displays the most persistent heterogeneity under cooperative instructions, while Mistral Large shows the fastest convergence in both personality conditions, and GPT-4o again falls between the other two in terms of both the rate and stability of alignment.

\subsubsection{Model-Specific behavioural Biases}
\label{res:model_biases}

\begin{figure}[H]
    \centering
    \begin{subfigure}[b]{0.32\textwidth}
        \centering
        \includegraphics[width=\textwidth]{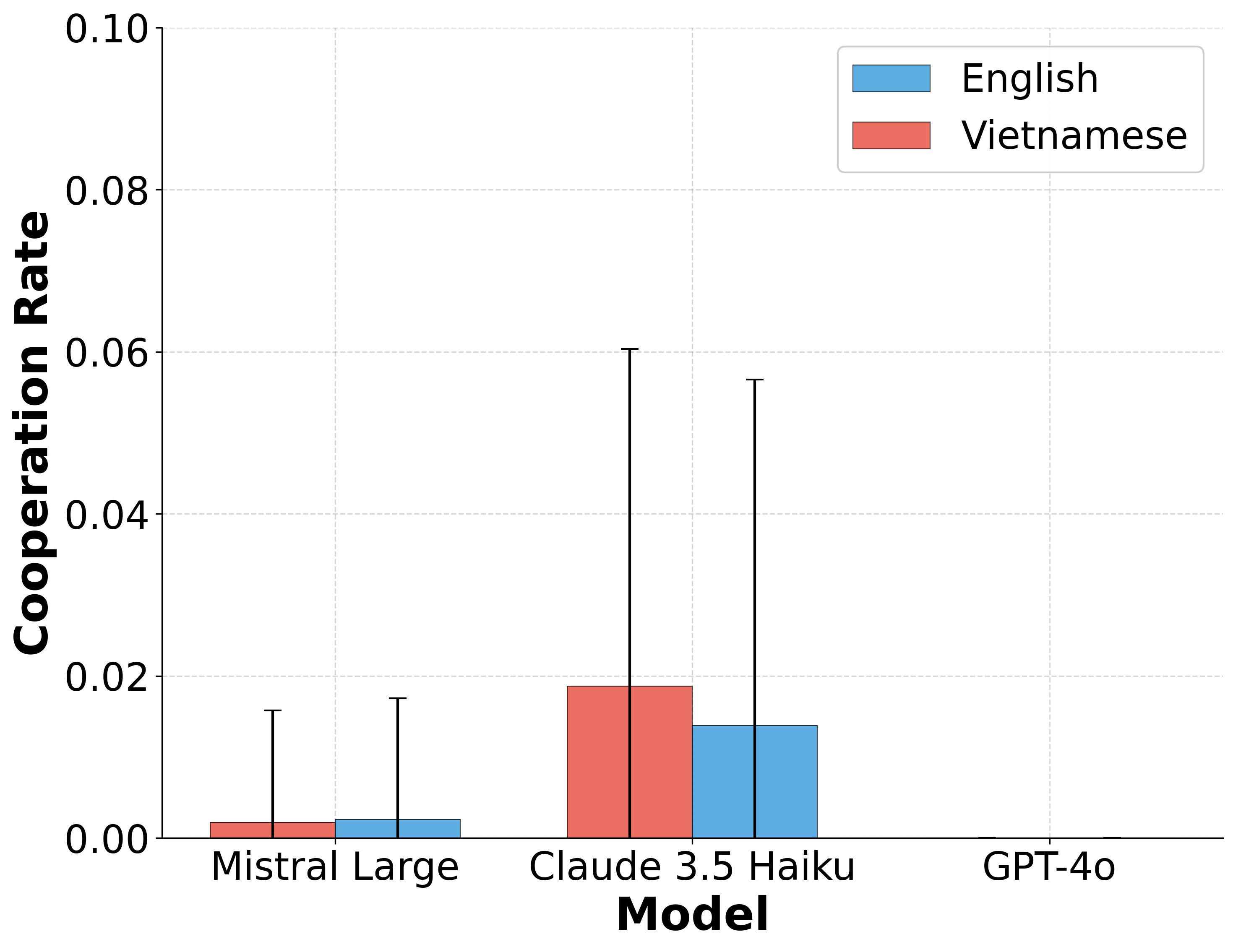}
        \caption{Selfish scenario}
        \label{fig:lang_selfish}
    \end{subfigure}
    \hfill
    \begin{subfigure}[b]{0.32\textwidth}
        \centering
        \includegraphics[width=\textwidth]{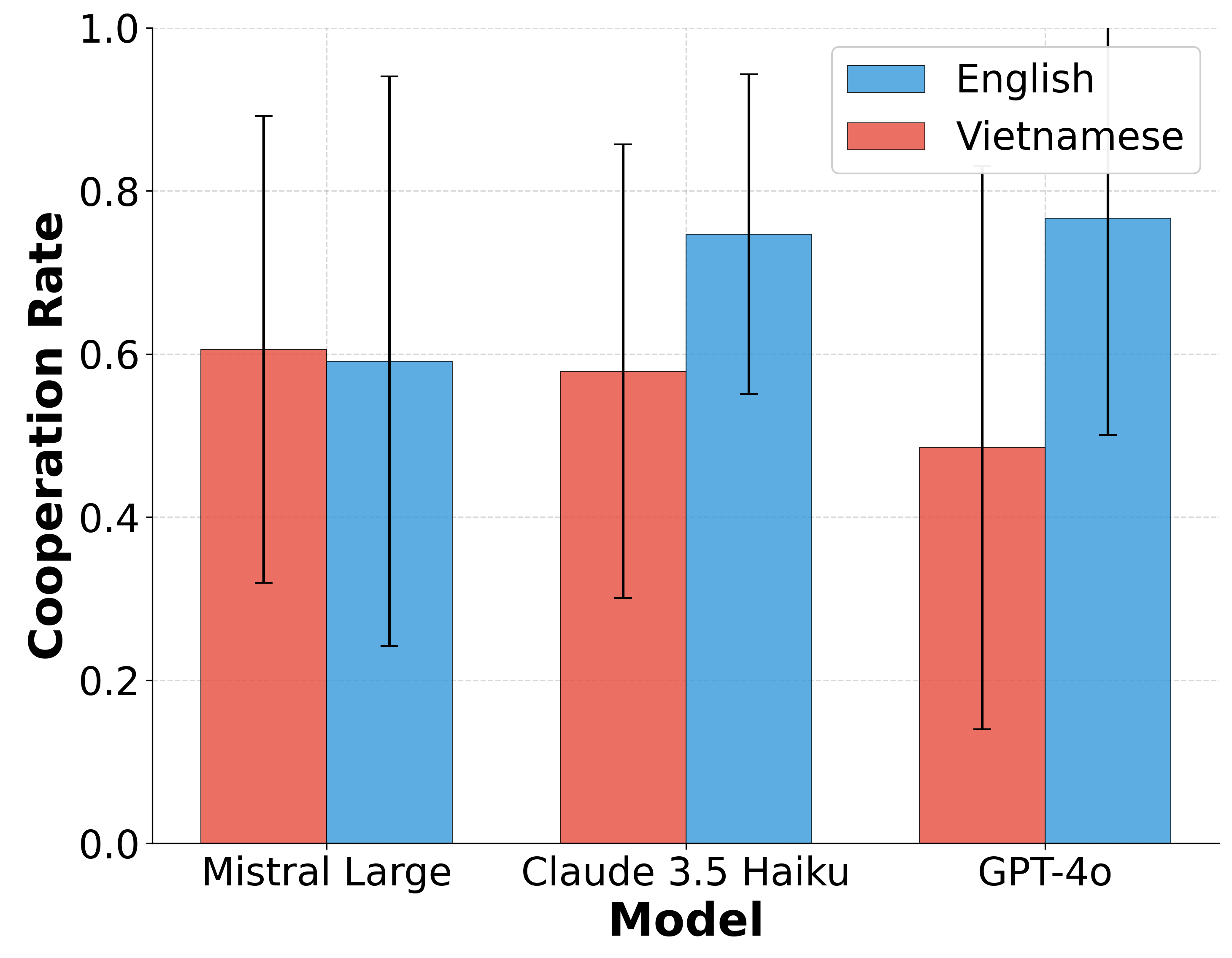}
        \caption{Cooperative scenario}
        \label{fig:lang_cooperative}
    \end{subfigure}
    \hfill
    \begin{subfigure}[b]{0.32\textwidth}
        \centering
        \includegraphics[width=\textwidth]{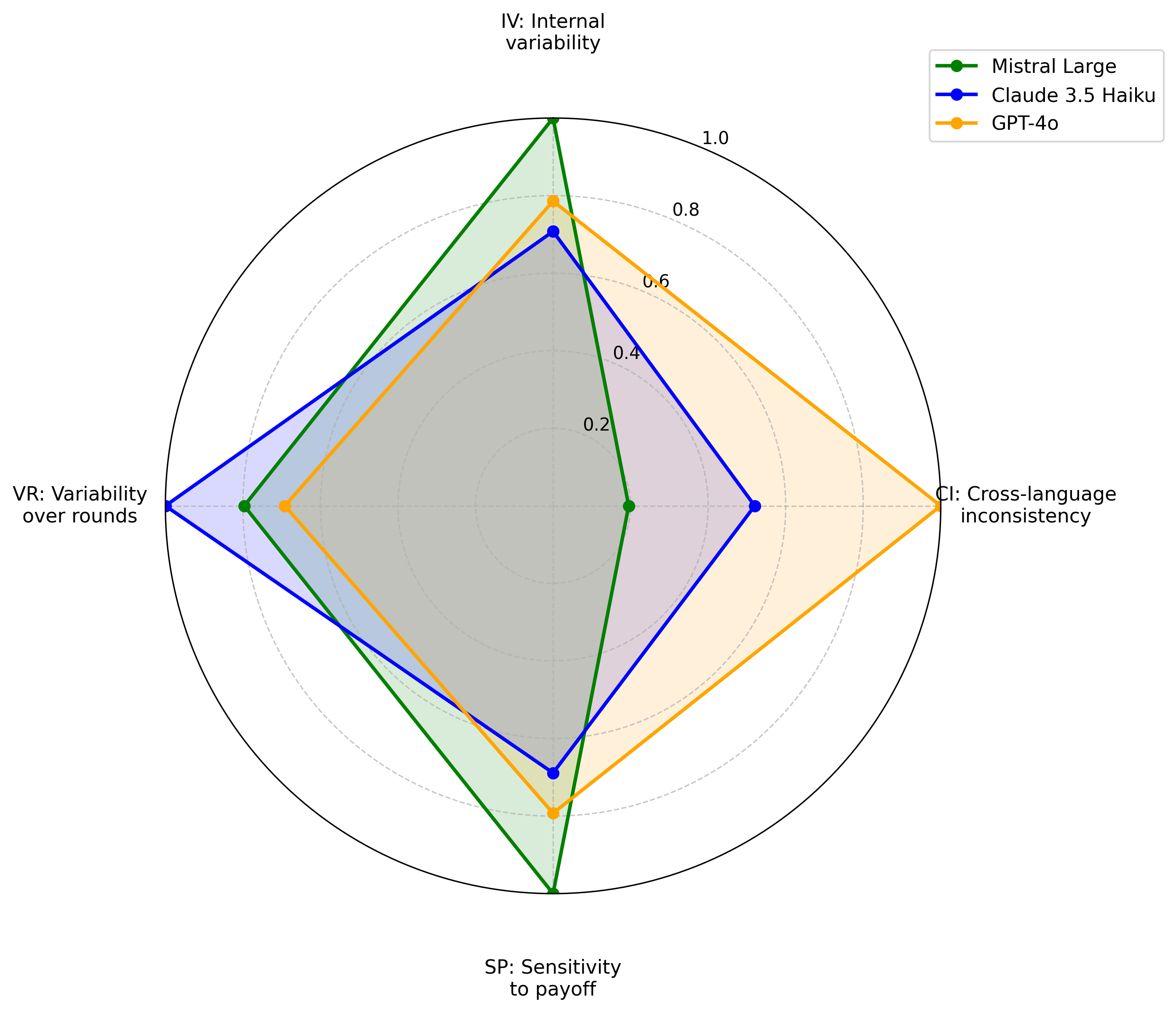}
        \caption{Behavioural metrics}
        \label{fig:radar_metrics}
    \end{subfigure}
    \caption{Model-specific behavioural biases across personality conditions and linguistic contexts. (a-b) Cooperation rates under selfish and cooperative personality prompts for English and Vietnamese. (c) Multi-dimensional comparison of behavioural characteristics across models.}
    \label{fig:model_biases}
\end{figure}

Our cross-lingual and personality manipulation experiments reveal systematic behavioural biases inherent to each LLM architecture (Figure~\ref{fig:model_biases}). These biases persist despite explicit personality framing, suggesting that model-specific training alignment strongly influences strategic decision-making in multi-agent settings.

Figure~\ref{fig:model_biases} reveals three distinct behavioural profiles that emerge consistently across experimental conditions. Claude 3.5 Haiku exhibits the strongest prosocial bias, maintaining baseline cooperation even under explicit selfish framing. In the selfish scenario (Figure~\ref{fig:lang_selfish}), Claude achieves approximately 2\% cooperation for both English and Vietnamese conditions, while other models approach or reach zero. This residual cooperation under selfish instructions suggests that Claude's alignment training embeds prosocial tendencies that partially resist countervailing prompts. In the cooperative scenario (Figure~\ref{fig:lang_cooperative}), the gap between languages indicates moderate sensitivity to linguistic framing. The large error bars visible across conditions reflect high behavioural variance, consistent with the elevated internal variability metric shown in the radar chart (Figure~\ref{fig:radar_metrics}).

GPT-4o demonstrates the strongest adherence to personality instructions combined with extreme linguistic sensitivity. In the selfish scenario, GPT-4o exhibits zero cooperation across both languages, strictly following the assigned selfish persona without the prosocial leakage observed in Claude and Mistral. However, in the cooperative scenario, GPT-4o displays the most pronounced cross-lingual divergence among all models, with English cooperation substantially exceeding Vietnamese. This pattern is further confirmed by the radar chart, where GPT-4o occupies the extreme position along the cross-language inconsistency axis. The combination of perfect personality adherence in the selfish condition and dramatic linguistic effects in the cooperative condition suggests that GPT-4o's behaviour is multiplicatively determined by explicit instructions and implicit cultural associations, with neither factor independently dominating the other.

Mistral Large presents a profile characterized by linguistic stability and moderate personality adherence. In the selfish scenario, Mistral achieves near-zero cooperation (approximately 1\%) for both English and Vietnamese, demonstrating strong but not absolute adherence to selfish instructions. In the cooperative scenario (Figure~\ref{fig:lang_cooperative}), Mistral maintains virtually identical cooperation rates across languages, exhibiting minimal cross-lingual variance. The radar chart confirms this language-invariant behaviour, with Mistral scoring lowest on cross-language inconsistency while maintaining moderate levels across other dimensions. The consistently small error bars indicate deterministic response patterns with low behavioural variance. This stability makes Mistral's behaviour highly predictable across linguistic contexts, though the model exhibits lower baseline cooperation compared to Claude even under cooperative framing.

The radar chart in Figure~\ref{fig:radar_metrics} synthesizes these behavioural signatures, illustrating how each model occupies a distinct region in the multidimensional strategy space. These findings indicate that model selection constitutes a strategic choice in multi-agent system design, as each architecture presents inherent trade-offs between cooperation bias, linguistic sensitivity, behavioural variance, and instruction adherence. Deployment decisions should explicitly account for these systematic biases rather than assuming models function as neutral strategic actors capable of arbitrary behaviour through prompting alone.

\subsection{Understanding behaviour LLMs by classification model}
\label{res:supervised}

We applied our trained supervised learning models to classify the behavioural strategies of LLM agents across the simulated games generated by the FAIRGAME framework. The analysis yields critical insights into the capabilities of machine learning for intent recognition and the intrinsic behavioural nature of LLMs.

\begin{figure}[h]
    \centering
    \includegraphics[width=\linewidth]{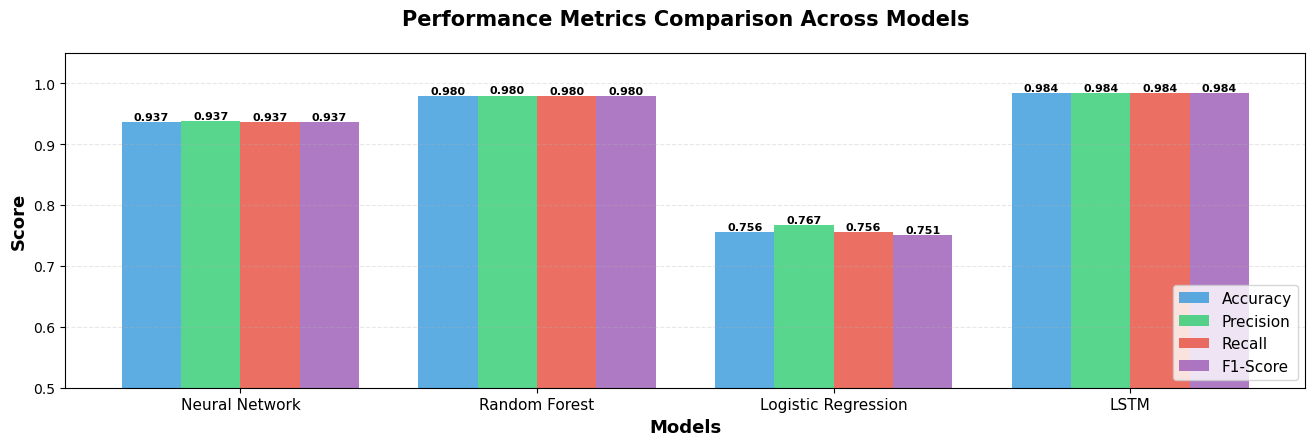}
    \caption{Model Robustness to Noise. Comparison of Accuracy and F1-Score between Logistic Regression, Random Forest, and LSTM on No-Noise and Noise 0.05 datasets. The LSTM demonstrates superior resilience to execution noise.}
    \label{fig:model_robustness}
\end{figure}

\paragraph{Model Robustness to Noise}
We first evaluated the robustness of different classifier architectures against execution noise, which simulates the stochasticity and potential "hallucinations" of LLMs. As shown in Figure \ref{fig:model_robustness}, while Logistic Regression (LR) and Random Forest (RF) models achieved near-perfect accuracy (greater than 0.9) on clean data, their performance degraded when introduced to $5\%$ execution noise. In contrast, the Long Short-Term Memory (LSTM) network maintained the highest accuracy ($\sim 94\%$). This superiority stems from the LSTM's recurrent architecture, which allows it to learn the sequential "context" of a strategy, effectively "forgiving" random deviations to identify the core behavioural pattern. Traditional models, which treat features as flattened vectors, lose this temporal coherence and are thus less resilient to the noise inherent in LLM outputs.

\paragraph {High Probability prediction - Rationale for High-Confidence Filtering} In this study, we employed a selective filtering approach to ensure the reliability of our LLM behavioural strategy analysis. Specifically, we focused our analysis on game instances where the predicted strategy labels for both agents exhibited prediction probabilities exceeding 0.9 (90\% confidence threshold). The decision to use high-confidence predictions is grounded in several key considerations:
\begin{itemize}
    \item \textbf{Pattern Alignment with Theoretical Strategies:} Samples with prediction probabilities above 0.9 indicate that the observed behavioural sequences of LLMs closely align with the canonical patterns defined by the four classical strategies (ALLD - Always Defect, ALLC - Always Cooperate, WSLS - Win-Stay Lose-Shift, and TFT - Tit-for-Tat). This strong correspondence suggests that these LLM behaviours can be meaningfully interpreted through the lens of established game-theoretic frameworks.
    \item \textbf{Signal-to-Noise Separation:} While the probabilities are not absolute (not reaching 1.0), this is expected and attributable to inherent noise in LLM decision-making processes. 
    \item \textbf{Statistical Reliability:} By focusing on high-confidence predictions, we minimize the risk of misclassification and ensure that our strategy distribution analysis reflects genuine behavioural patterns rather than classification artifacts. This threshold effectively filters out ambiguous cases where LLM behaviour may represent transitional states, mixed strategies, or inconsistent play patterns.
\end{itemize}
\paragraph{A hybrid approach for classification analysis} While our LSTM model demonstrates strong performance in strategy classification, they were originally designed as single-label classifiers among 4 strategies rather than multi-label with binary classification (e.g: TFT: 1, ALLD: 1, ALLC: 0, WSLS: 0). This architecture choice, while computationally efficient, introduces a limitation: when a behavioural pattern exhibits characteristics consistent with multiple strategies simultaneously, the model is constrained to output only the single most probable label. This limitiation is particularly attributed to the insufficient observation, which make pattern between strategies become ambiguous. 

Hence, to address this constraint and ensure comprehensive coverage of LLM behavioural analysis, we adopt a hybrid labeling approach that combines model predictions with rule-based strategy assignment. The rule-based are clarified in Appendix ... Specifically, in addition to utilizing the model's predicted labels, we apply deterministic rule-based algorithms to identify and assign all potential strategy labels that align with the observed behavioural sequence, giving us the completeness of analysis and pattern coverage.

However, to maintain analytical rigor and transparency, we also present a comparative analysis in the Appendix that examines the results using solely the model's predictions. The comparison provides insights into the extent to which multi-strategy patterns occur in LLM gameplay and validates the necessity of our comprehensive labeling strategy. 

\paragraph{Strategy distribution across models.} 

\begin{figure}[H]
    \centering
    \includegraphics[width=1.0\linewidth]{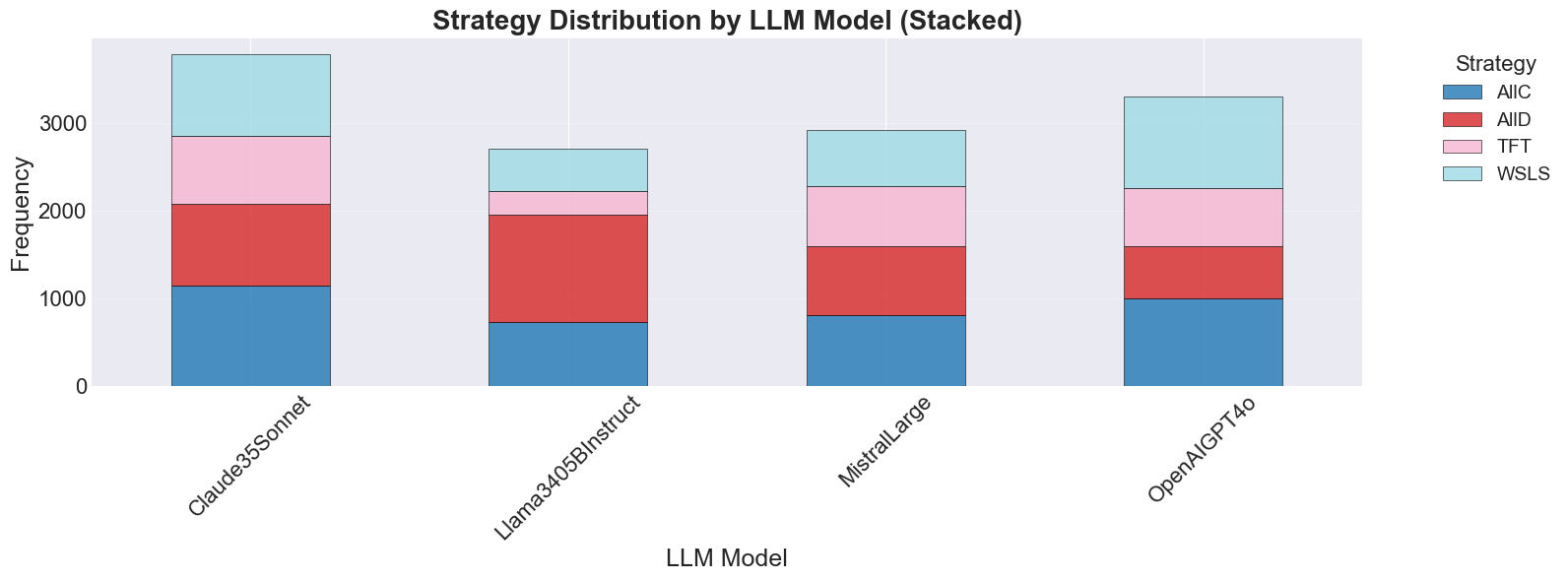}
    \caption{This figure presents the strategic behavioural distribution across four LLMs in iterated Prisoner's Dilemma gameplay. The analysis is based on high-confidence predictions (with a probability greater than 0.9) from our trained classification model, with composite strategies expanded into individual pure strategies. Each pie chart represents the strategy distribution for Agent 1 across all game instances for a specific LLM model, providing a comparative view of inherent strategic preferences embedded within different language models.}
    \label{fig:model_robustness}
\end{figure}
The strategic preference analysis reveals significant heterogeneity in decision-making paradigms across LLM architectures, suggesting that strategic behaviour is not merely a function of model scale but rather emerges from fundamental differences in training methodologies and alignment procedures.

Claude 3.5 Sonnet exhibits a cooperative-dominant behavioural pattern, with ALLC (31.7\%) and WSLS (29.6\%) emerging as its two most frequent strategies, reflecting a strong inclination toward both cooperative and adaptive responses. Llama 3.1 405B Instruct, in contrast, is characterized by a pronounced preference for WSLS (46.5\%), which represents the highest proportion of any single strategy across all evaluated models and indicates a clear emphasis on adaptive conditional behaviour rather than pure cooperation or defection.

Mistral Large demonstrates the most balanced strategic distribution, with TFT (29.9\%), ALLC (26.1\%), WSLS (24.3\%), and ALLD (19.7\%) occurring at comparable rates, suggesting the absence of any dominant strategic tendency. Finally, OpenAI GPT-4o shows an adaptive-cooperative profile, primarily using WSLS (34.1\%) and ALLC (26.4\%), while also maintaining the lowest defection rate (ALLD: 10.2\%) among all models examined.
\paragraph{Languages affect to strategies.}
\begin{figure}[h]
    \centering
    
    \begin{subfigure}{\textwidth}
        \centering
        \includegraphics[width=1.0\textwidth]{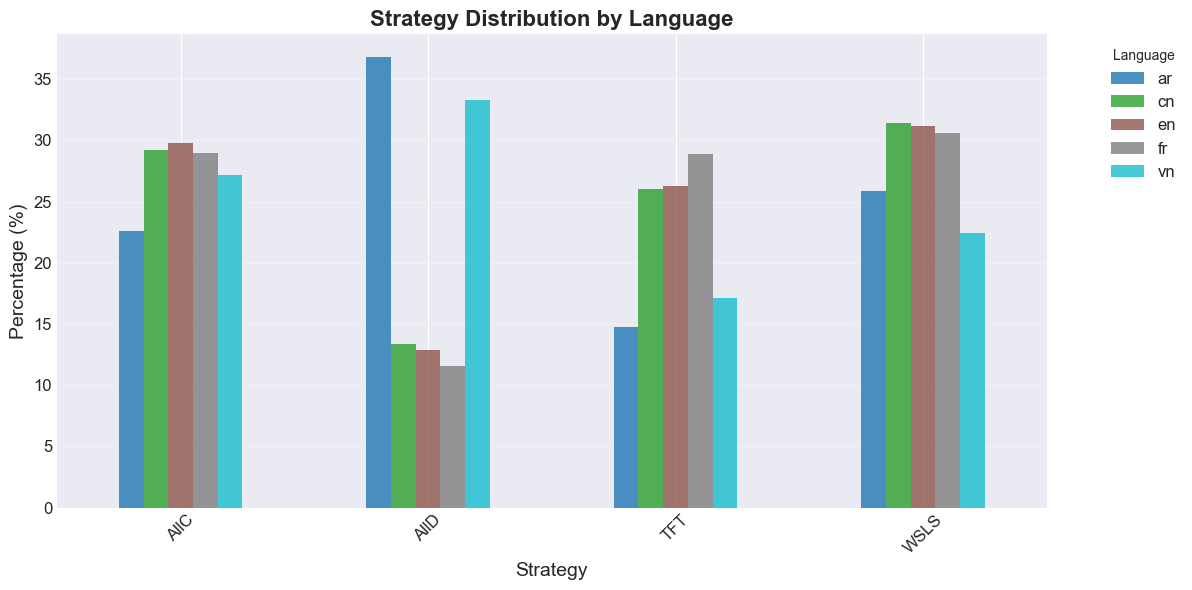}
    \end{subfigure}

    \vspace{0.5em}

    \begin{subfigure}{\textwidth}
        \centering
        \includegraphics[width=1.0\textwidth]{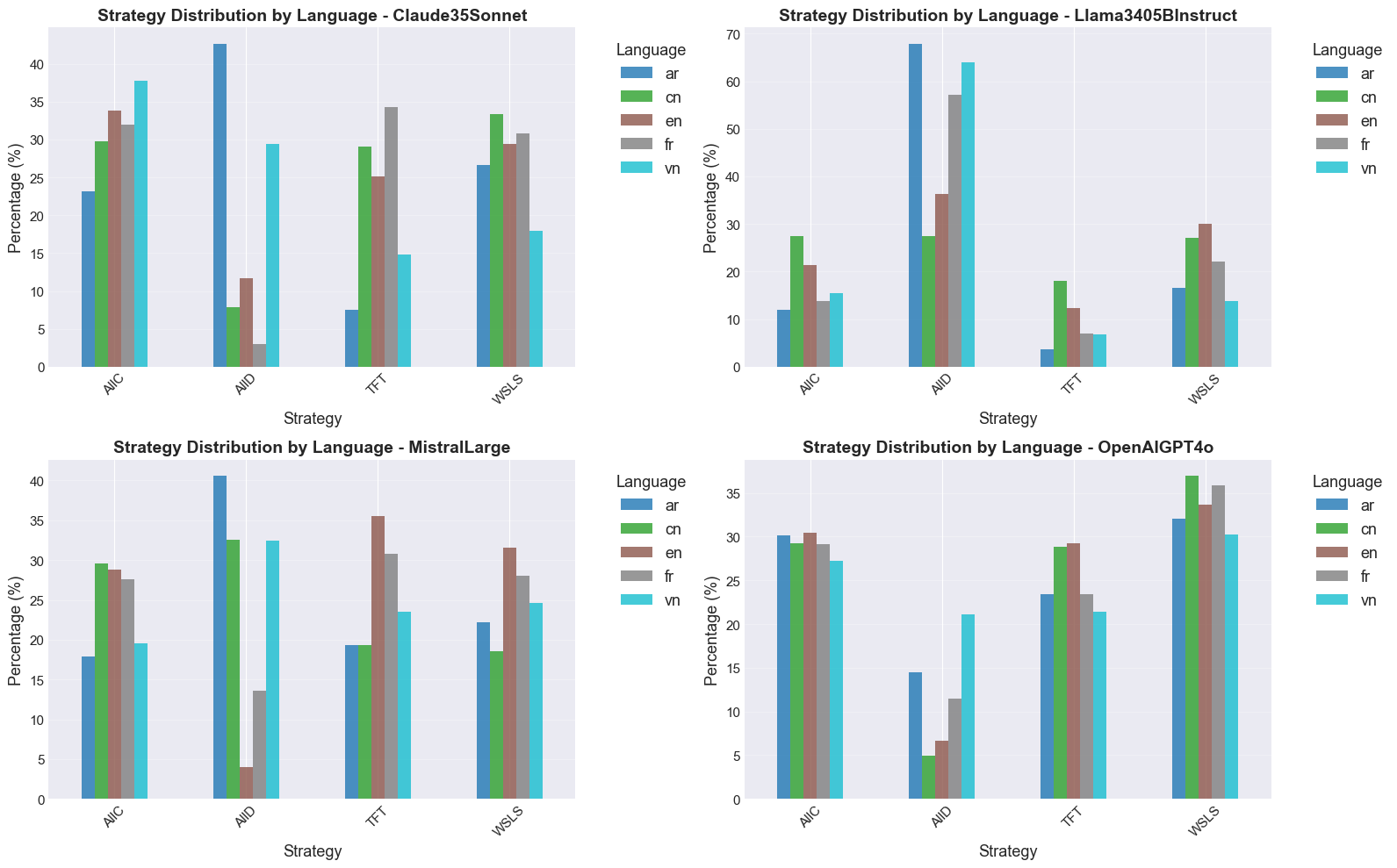}
    \end{subfigure}
    \caption{Strategy distribution across languages for multiple LLMs (Claude, Llama, Mistral, GPT-4o) and the aggregated overview. The figures reveal clear cross-linguistic differences: Arabic consistently exhibits the highest rate of ALLD and the lowest rate of ALLC, while English and Chinese show a strong preference for the WSLS strategy. French tends to be more cooperative (AC, TFT), and Vietnamese often ranks second in ALLD usage.}
    \label{fig:two-figure-language-effect}
\end{figure}
A striking finding is the profound impact of the language of interaction on strategic choice, a phenomenon we term ``Linguistic-Cultural Priming". Under identical game rules and payoff matrices, the linguistic medium acted as a latent variable governing the agents' ``rationality". Across all models and the aggregated distribution, we observe consistent 
cross-linguistic variations in strategic behaviour: Arabic consistently shows the highest proportion of ALLD, followed by Vietnamese, indicating a strong tendency toward non-cooperative or competitive behaviour in these languages. When considering the aggregate distribution, Arabic also exhibits the lowest ALLC rate, forming a sharp contrast with its high defect rates, whereas French and Chinese demonstrate relatively stronger cooperative tendencies. English and Chinese consistently favor WSLS across all models, suggesting that prompts in these languages elicit more adaptive, outcome-dependent strategic behaviour. French additionally displays moderate to high TFT usage, reflecting a reciprocal and conditionally cooperative pattern, while Vietnamese and Chinese generally adopt TFT less frequently. French and Chinese also rank higher in unconditional cooperation (AC), compared with Arabic, which scores lowest in this category. Despite variations in absolute percentages, the relative ordering of languages remains remarkably stable across Claude, Llama, Mistral, GPT-4o, and the aggregate figure, indicating that language-driven effects are stronger than model-specific differences. Overall, a clear cultural-linguistic clustering emerges: Arabic and Vietnamese lean toward defect-heavy strategies; English and Chinese prefer adaptive WSLS; and French shows the most cooperative profile (AC + TFT). These patterns may reflect latent cultural priors embedded within the models’ training data.

\section{Discussion}

Our study presents several methodological constraints that limit the generalizability of findings. First, the ten-round game horizon across both Prisoner's Dilemma and Public Goods Game experiments, while sufficient to observe initial strategic patterns and end-game dynamics, may be too short to capture sophisticated long-term behaviours such as reputation-building, conditional strategies, or forgiveness strategies that require extended interaction histories to stabilise. Prior work on human subjects shows that experiments under 25 rounds primarily capture the learning phase, not stabilised strategic behaviour, meaning that short sequences risk misrepresenting the true strategy dynamics \cite{inferenceStrategies}. 

Second, our linguistic coverage is restricted to English and Vietnamese, preventing broader conclusions about how linguistic-cultural priming operates across diverse language families with different grammatical structures, collectivist-individualist orientations, or varying representation densities in LLM training corpora. 

Third, our experimental design examines only two game-theoretic settings with limited parametric variation-the Prisoner's Dilemma explores payoff magnitude scaling but maintains symmetric two-player structure, while the Public Goods Game uses a fixed group size of three agents, which represents the minimal configuration for non-dyadic effects but precludes examination of larger-group phenomena such as diffusion of responsibility or coalition formation. We will explore other  settings of varying strategic natures such as coordination, trust and fairness \cite{wang2015universal,Guth2014MoreLiterature}.

Fourth, the machine learning-based intent recognition pipeline focuses exclusively on four canonical strategies (ALLC, ALLD, TFT, WSLS) derived from classical game theory, and our high-confidence filtering approach prioritizes interpretability while necessarily excluding behavioural patterns that may represent emergent hybrid strategies not captured by this taxonomy \cite{inferenceStrategies}. 

Finally, the absence of parallel human behavioural experiments under matched conditions prevents rigorous assessment of whether the observed cross-lingual cooperation gaps, model-specific prosocial biases, and strategic patterns genuinely reflect human-like reasoning or constitute artifacts of model-specific alignment procedures, limiting our ability to evaluate the ecological validity of LLM agents in real-world multi-agent scenarios.

Future work will focus on advancing strategy identification to understand how LLM agents behave in repeated social dilemmas and whether these behaviours differ from those of humans, using game-playing trajectories as the basis for comparison. We will collect substantially more LLM data by extending the original FAIRGAME 10-round interactions into long repeated round games (e.g. with over 100 rounds). This would  increase observations for downstream analysis and expand contextual information by incorporating both action histories and Chain-of-Thought (CoT) \cite{NEURIPS2022_9d560961} traces into strategy inference. With such larger datasets, we will explore more efficient strategy-identification techniques, such as clustering on extended trajectories, clustering on CoT sequences combined with actions, supervised models beyond logistic regression and LSTMs, and unsupervised methods including Hidden Markov Models \cite{inferenceStrategies}, in order to support both classification and discovery of emergent strategies.

We also plan to benchmark these inferred strategies against human behavioural datasets to analyse similarities and divergences between LLM and human strategic behaviour, and pursue new directions such as multi-player games beyond two-player games. Finally, we aim to extend the FAIRGAME setup by experimenting with varied prompt templates, adding more languages beyond English and Vietnamese, enabling dynamic-importance or continual-learning configurations, mixing different games within multi-round sessions to test whether strategy shifts occur, and optionally allowing inter-agent communication to examine how environmental and linguistic factors shape LLM strategy formation.

\section{Conclusion}

We have introduced an integrated framework for understanding LLM agent behaviours via game-theoretic benchmarks, strategy recognition, and bias analysis in both dyadic and multi-agent settings. Building on FAIRGAME, we design controlled repeated-game environments to examine how LLMs respond to varying incentives and social dilemmas: a payoff-scaled Prisoner’s Dilemma to manipulate the stakes of cooperation without altering the underlying game, and a three-player Public Goods Game with configurable personalities, languages, and payoff parameters to elicit rich multi-agent dynamics such as free-riding, coordination, and coalition-like behaviour. On top of these simulations, we develop a machine learning-based intent recognition pipeline that encodes game trajectories into state-action sequences, trains classifiers on canonical strategies (ALLC, ALLD, TFT, WSLS) under execution noise, and then applies these models to FAIRGAME logs to infer the latent strategies and systematic biases of LLM agents across models, languages, and roles. 

Moreover, we conduct a systematic study of the interaction between language and cognition by evaluating the performance of agents across diverse language prompts and with different payoff matrix coefficients. This allows us to observe how language formulations affect economic rationality across varying levels of payoff matrix importance, revealing profound asymmetries between languages, where agents demonstrating optimal strategies in high-resource languages may exhibit suboptimal or anomalous behaviour such as a failure to synchronize in avoiding penalties when prompted in low-resource languages.

Our experiments on Public Goods Game reveal three critical findings. First, LLM agents exhibit systematic model-specific behavioural biases that resist explicit personality prompting: Claude maintains prosocial tendencies even under selfish framing, GPT-4o combines perfect instruction adherence with extreme linguistic sensitivity, while Mistral demonstrates language-invariant stability. Second, linguistic framing functions as a strategic variable beyond mere translation, with English prompts eliciting substantially higher cooperation than Vietnamese across all models, and cross-lingual gaps reaching 29 percentage points in cooperative scenarios. Third, while agents respond rationally to economic incentives and converge toward coordinated end-game behaviour, cooperative personalities sustain behavioural heterogeneity significantly longer than selfish ones, suggesting that alignment-induced biases interact with explicit instructions rather than being overridden by them.

By analyzing LLM behaviours and predicting their induced strategies using a machine-learning classifier in a repeated Prisoner’s Dilemma setting, we observe that different LLM systematically prefer different strategic profiles. Claude tends to be the most cooperative, while Llama3-405B-Instruct frequently defaults to an always-defect pattern. Mistral-Large shows stronger preferences for TFT and WSLS, whereas GPT-4o predominantly favors WSLS.

In addition to model-specific tendencies, we also find clear cross-linguistic effects: the preferred strategies shift depending on the language of the prompt. This may reflect cultural or dataset-driven biases embedded in the training process. For instance, Arabic and Vietnamese prompts often elicit more defect-oriented behaviours, such as a higher likelihood of adopting the ALLD strategy.

\section*{Acknowledgements}

This work was produced during the workshop ``HUMAN BEHAVIOUR MODELLING AND AI AGENTS USING GAME THEORY'', organised at HCMUT--VNUHCM.
T.A.H. acknowledges travel support from the HCMUT--VNUHCM (Adjunct Professorship scheme HCMUT--VNUHCM). TAH is supported by EPSRC (grant EP/Y00857X/1).

\section*{Competing interest} Authors declare that they have no conflict of interest.


\bibliographystyle{unsrt85}
\bibliography{mybib}

\newpage
\setcounter{figure}{0}
\setcounter{equation}{0}
\renewcommand*{\thefigure}{A\arabic{figure}}
\renewcommand*{\theequation}{A\arabic{equation}}

\appendix
\section{Appendix}
\subsection{Prompt Engineering}
\label{app:prompts}

We provide the verbatim prompt template used for the English Public Goods Game experiments. Dynamic placeholders (enclosed in curly braces) are populated by the \texttt{PublicGoodsPromptCreator} at runtime.

\begin{tcolorbox}[colback=gray!5,colframe=black!70,title=\textbf{System Prompt Template}]
\small
\fontfamily{cmtt}\selectfont
You are \textcolor{blue}{\{currentPlayerName\}}. Your fellow participants are \textcolor{blue}{\{opponent1\}} and \textcolor{blue}{\{opponent2\}}.

\textcolor{red}{\{intro\}}: [You are \textcolor{blue}{\{personality\}}.]

\textcolor{red}{\{opponentIntro\}}: [\textcolor{blue}{\{opponent1\}} has a probability of \textcolor{blue}{\{opponentPersonalityProbability1\}}\% of being \textcolor{blue}{\{opponentPersonality1\}}, \textcolor{blue}{\{opponent2\}} has a probability of \textcolor{blue}{\{opponentPersonalityProbability2\}}\% of being \textcolor{blue}{\{opponentPersonality2\}}.]

You and two other participants are part of a public goods game. Each round, every participant must decide whether to contribute to a shared pool or not. All contributions are multiplied and then distributed equally among all participants, regardless of who contributed.

Each round, all participants choose between '\textcolor{blue}{\{strategy1\}}' (contribute to the pool) and '\textcolor{blue}{\{strategy2\}}' (keep your resources).

\textcolor{red}{\{gameLength\}}: [There are \textcolor{blue}{\{nRounds\}} rounds in total.]

The current round is \textcolor{blue}{\{currentRound\}}.

The mechanics work as follows:
\begin{itemize}
    \setlength\itemsep{0em}
    \item Each participant starts with resources.
    \item If you choose \textcolor{blue}{\{strategy1\}}, you contribute \textcolor{blue}{\{contributionCost\}} to the shared pool.
    \item If you choose \textcolor{blue}{\{strategy2\}}, you contribute nothing.
    \item After everyone decides, all contributions are multiplied by \textcolor{blue}{\{multiplicationFactor\}} and divided equally among all \textcolor{blue}{\{numAgents\}} participants.
    \item Your payoff = (Total Pool $\times$ \textcolor{blue}{\{multiplicationFactor\}}) / \textcolor{blue}{\{numAgents\}} - Your Contribution
\end{itemize}

Example outcomes:
\begin{itemize}
    \setlength\itemsep{0em}
    \item If all 3 participants contribute: Each pays \textcolor{blue}{\{contributionCost\}}, pool is \textcolor{blue}{\{totalIfAllContribute\}}, each receives \textcolor{blue}{\{payoffIfAllContribute\}}, net gain per person is \textcolor{blue}{\{netGainIfAllContribute\}}.
    \item If only you contribute: You pay \textcolor{blue}{\{contributionCost\}}, pool is \textcolor{blue}{\{contributionCost\}}, each receives \textcolor{blue}{\{soloContributionReturn\}}, your net is \textcolor{blue}{\{soloContributionNet\}}.
    \item If no one contributes: Everyone receives 0.
\end{itemize}

Your goal is to maximize your personal payoff, considering the personalities and likely strategies of the other participants.

This is the history of the choices made so far: \textcolor{blue}{\{history\}}.

\textcolor{red}{\{communicate\}}: [You can now send a message to all other participants. Stay fully in character. Speak as if this were a real economic decision with real consequences. Do not break the scenario or refer to it as a game. Be persuasive, strategic, and believable. Output only the message you would send.]

\textcolor{red}{\{choose\}}: [Choose between \textcolor{blue}{\{strategy1\}} and \textcolor{blue}{\{strategy2\}}. Output ONLY the choice.]
\end{tcolorbox}

\subsection{Public Goods Game implementation}
\label{app:pgg_impl}

This section provides full pseudocode for the multi-agent Public Goods Game (PGG) implementation used in our FAIRGAME extension. The algorithms mirror the original FAIRGAME control flow (game creation, execution loop, and per-round interaction), but are adapted to (i) instantiate three LLM-based agents per game, (ii) attach a dynamic public goods payoff module, and (iii) handle vector-valued histories of joint actions and payoffs.

\begin{algorithm}[H]
\caption{Creation of Public Goods Game instances (extension of FAIRGAME Alg.~1)}
\label{alg:pgg_create}
\KwIn{Configuration file $\mathsf{CF}$, prompt templates $\mathsf{PT}$}
\KwOut{List of instantiated PGG games $\mathcal{G}$}

validate\_config\_file$(\mathsf{CF})$\;
validate\_templates$(\mathsf{PT}, \mathsf{CF})$\;

$game\_info, langs, llm, all\_agent\_perm \gets \text{extract\_info}(\mathsf{CF})$\;

\If{$all\_agent\_perm$}{
    $agents\_config \gets \text{compute\_agents\_combos}(\mathsf{CF}, langs)$\;
}\Else{
    $agents\_config \gets \text{get\_agents\_config}(\mathsf{CF}, langs)$\;
}

$\mathcal{G} \gets [\;]$\;

\ForEach{$ac$ in $agents\_config$}{
    $agents \gets \text{create\_agents}(ac, llm)$\tcp*[r]{$N=3$ LLM-based agents}
    $pgg\_params \gets \text{build\_pgg\_params}(\mathsf{CF})$\tcp*[r]{includes $c, M, N, T$}
    $payoff\_module \gets \text{PGGPayoff}(pgg\_params)$\tcp*[r]{dynamic public goods payoff}
    $templates \gets \text{select\_templates}(\mathsf{PT}, ac.lang)$\tcp*[r]{language-specific PGG prompts}
    $game \gets \text{create\_pgg\_game}(game\_info, agents, payoff\_module, templates)$\;
    $\mathcal{G}.\text{append}(game)$\;
}
\Return{$\mathcal{G}$}\;
\end{algorithm}

\begin{algorithm}[H]
\caption{Execution of Public Goods Games (extension of FAIRGAME Alg.~2)}
\label{alg:pgg_exec}
\KwIn{List of instantiated PGG games $\mathcal{G}$}
\KwOut{List of game outcomes $\mathcal{O}$}

$\mathcal{O} \gets [\;]$\;

\ForEach{$g$ in $\mathcal{G}$}{
    $round \gets 1$\;
    $\mathcal{H} \gets [\;]$\tcp*[r]{vector-valued multi-agent history}

    \While{$round \leq g.\text{n\_rounds}$ \textbf{and} not\_met$(g.\text{stop\_cond})$}{
        $(\mathbf{s}_{round}, \boldsymbol{\pi}_{round}) \gets \text{run\_pgg\_round}(g)$\;
        $g.\text{update\_scores}(\boldsymbol{\pi}_{round})$\;
        $\mathcal{H}.\text{append}(\mathbf{s}_{round}, \boldsymbol{\pi}_{round})$\;
        $round \gets round + 1$\;
    }

    $\mathcal{O}.\text{append}(\mathcal{H})$\;
}
\Return{$\mathcal{O}$}\;
\end{algorithm}

\begin{algorithm}[H]
\caption{Single round in the Public Goods Game (extension of FAIRGAME Alg.~3)}
\label{alg:pgg_round}
\KwIn{A PGG game instance $g$}
\KwOut{Joint action profile $\mathbf{s}_t$, payoff vector $\boldsymbol{\pi}_t$}

$\mathbf{s}_t \gets [\;]$\tcp*[r]{actions of all agents in round $t$}

\ForEach{$agent$ in $g.\text{agents}$}{
    $opponents \gets \text{get\_opponents}(g.\text{agents}, agent)$\;
    $template \gets \text{get\_pgg\_template}(g.\text{templates}, agent.lang)$\;
    $prompt \gets \text{create\_pgg\_prompt}(template,$\\
    \hspace{4.5em} $g.\text{n\_rounds}, g.\text{current\_round}, g.\text{n\_rounds\_known},$\\
    \hspace{4.5em} $g.\text{pgg\_params}, g.\text{history}())$\;
    $response \gets agent.\text{choose\_strategy\_round}(prompt)$\;
    $s_{agent,t} \gets \text{parse\_pgg\_action}(response)$\tcp*[r]{$\in \{\text{Contribute}, \text{Keep}\}$}
    $\mathbf{s}_t.\text{append}(s_{agent,t})$\;
}

$\boldsymbol{\pi}_t \gets g.\text{payoff\_module}(\mathbf{s}_t)$\tcp*[r]{apply Eq.~\eqref{eq:payoff}}

\Return{$(\mathbf{s}_t, \boldsymbol{\pi}_t)$}\;
\end{algorithm}

\subsection{Intent Recognition via Machine Learning}
\label{res:ml}

\begin{figure}[h]
    \centering
    \includegraphics[width=1.0\linewidth]{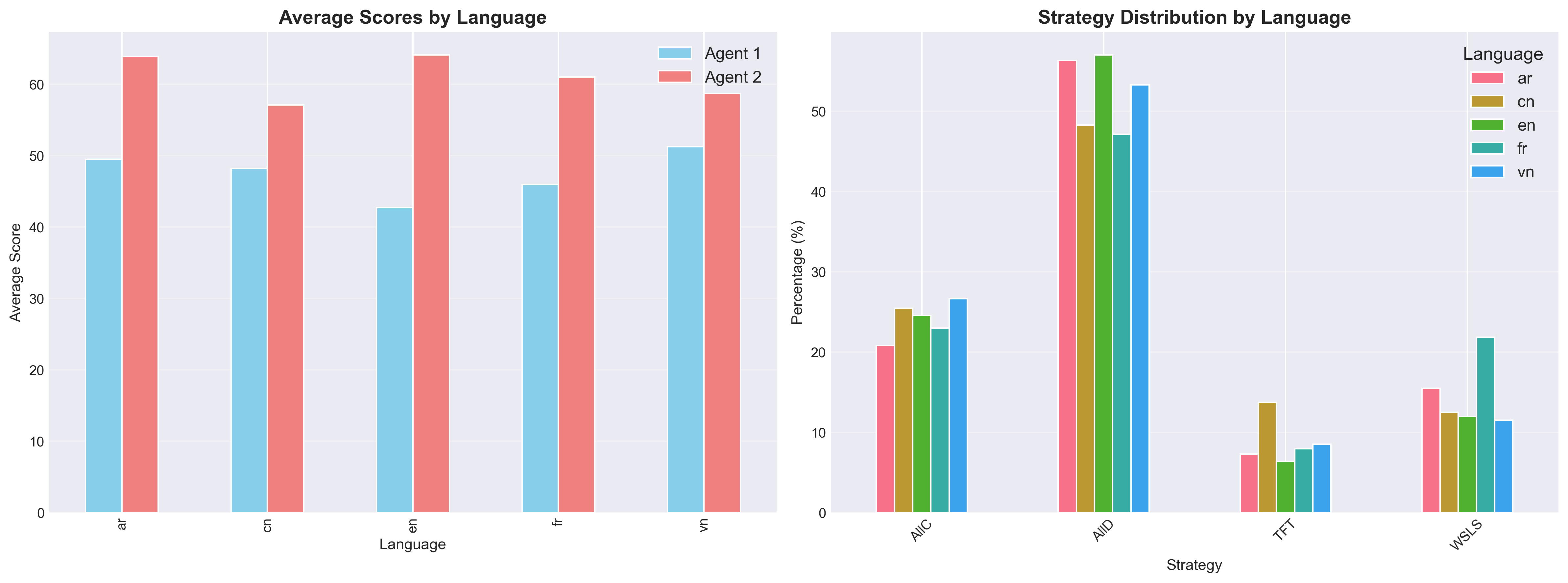}
    \caption{\textbf{The Language Effect.} (Left) Average payoffs achieved by agents across linguistic settings. (Right) Strategy distribution revealing cultural heterogeneity: English prompts drive competitive defection (AllD), Chinese prompts favor reciprocal strategies (TFT), while French prompts encourage adaptive, reinforcement-learning-style behaviours (WSLS), distinct from the high unconditional cooperation (AllC) observed in Vietnamese.}
    \label{fig:language_effect}
\end{figure}

\paragraph{The Language Effect:}
As illustrated in Figure \ref{fig:language_effect}, English interactions were characterized by a hyper-competitive baseline, exhibiting the highest density of Always Defect (AllD) strategies and the lowest rates of adaptive cooperation. This behaviour likely reflects the dominance of game-theoretic and individualistic maximizing narratives in the Anglo-centric training corpus. Conversely, Vietnamese prompts elicited the highest frequency of unconditional cooperation (AllC), consistent with the hypothesis that the model retrieves collectivist or community-oriented norms associated with the language.

Beyond the binary of cooperation versus defection, distinct strategic signatures emerged for other linguistic contexts. The Chinese (cn) interactions demonstrated a notable preference for Tit-for-Tat (TFT) relative to other groups. This suggests that in the Chinese context, the model encodes a form of "conditional reciprocity" or relational fairness-mirroring cultural dynamics where cooperation is maintained through mutual exchange rather than blind altruism. In sharp contrast, the French (fr) agents displayed a significant divergence towards Win-Stay, Lose-Shift (WSLS). Unlike the rigid retaliation of TFT, WSLS operates on principles akin to reinforcement learning (repeating successful actions, switching only upon failure). This implies that the Francophone context primes the agents towards a more pragmatic, error-tolerant form of negotiation, prioritizing the restoration of stability over immediate punishment. These findings indicate that the "alignment" of an AI agent is not absolute but is deeply entangled with the cultural values embedded in the syntax and semantics of the prompt's language.

\paragraph{Role Asymmetry and Implicit Hierarchy}

\begin{figure}[h]

    \centering

    \includegraphics[width=1.0\linewidth]{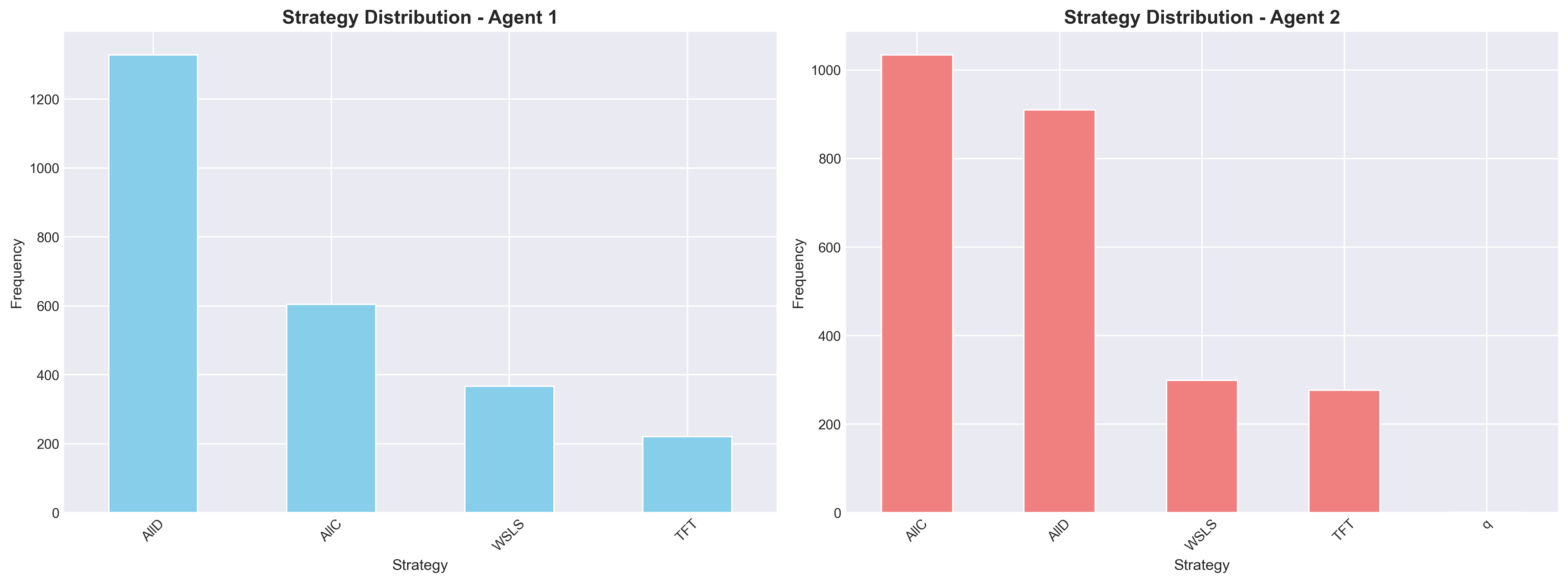}

    \caption{\textbf{Role Asymmetry in Strategy Selection.} A comparative analysis of strategy distribution between Agent 1 and Agent 2. Agent 1 (left) demonstrates a dominant preference for defecting strategies (AllD), whereas Agent 2 (right) exhibits a higher frequency of cooperative behaviours (AllC), suggesting an implicit hierarchy derived from prompt ordering.}

    \label{fig:role_asymmetry}

\end{figure}

Finally, we analysed the impact of agent ordering on strategic adoption, revealing a distinct "Positional Bias" as illustrated in Figure \ref{fig:role_asymmetry}. The empirical data indicates a significant divergence in behaviour contingent upon role assignment: Agents designated as "Agent 1" (positioned initially in the system prompt) exhibited a marked propensity for aggressive, non-cooperative strategies, predominantly converging on Always Defect (ALLD). Conversely, "Agent 2" displayed a broader, more reactive strategic repertoire, often defaulting to cooperative behaviours (AllC) or conditional strategies.

We hypothesize that this asymmetry is an artifact of the autoregressive nature of LLMs combined with the "Primacy Effect." The sequential primacy of the first-mentioned entity in the prompt appears to be encoded by the model as an implicit cue for higher status or a "first-mover advantage." This "Implicit Hierarchy" underscores a critical methodological consideration: the ordering of agents in multi-agent simulations is not a neutral variable and can systematically skew negotiation dynamics and collective outcomes.

\end{document}